\documentclass[twocolumn,twocolappendix]{aastex631}

\usepackage{makecell}
\usepackage{array}
\usepackage{amsmath}
\usepackage{textcomp} 
\usepackage{multirow}
\usepackage{enumitem}

\newcommand\lya{Ly$\alpha$ }

\newcommand{\pc}{\, \text{pc}}
\newcommand{\Mpc}{\, \text{Mpc}}
\newcommand{\hinvmpc}{\,h^{-1}{\rm Mpc}}

\newcommand{\kms}{\,{\rm km\,s^{-1}}}
\newcommand{\kmsMpc}{\,{\rm km\,s^{-1}\,Mpc^{-1}}}

\newcommand{\ergs}{\,{\rm erg\,s^{-1}}}
\newcommand{\K}{\,{\rm K}}
\newcommand{\cubiccms}{\,{\rm cm^{3}\,s^{-1}}}

\begin{document}

\title{Measurements of quasar proximity zones with the \lya forest of DESI Y1 quasars}

\author[0000-0002-5959-9199]{Ryuichiro~Hada}
\email{rhada@asiaa.sinica.edu.tw}
\affiliation{Department of Astronomy, The Ohio State University, 4055 McPherson Laboratory, 140 W 18th Avenue, Columbus, OH 43210, USA}
\affiliation{Center for Cosmology and AstroParticle Physics, The Ohio State University, 191 West Woodruff Avenue, Columbus, OH 43210, USA}
\affiliation{Institute of Astronomy and Astrophysics, Academia Sinica, No. 1, Sec. 4, Roosevelt Rd, Taipei 106319, Taiwan}

\author[0000-0002-4279-4182]{Paul~Martini}
\affiliation{Department of Astronomy, The Ohio State University, 4055 McPherson Laboratory, 140 W 18th Avenue, Columbus, OH 43210, USA}
\affiliation{Center for Cosmology and AstroParticle Physics, The Ohio State University, 191 West Woodruff Avenue, Columbus, OH 43210, USA}
\affiliation{Department of Physics, The Ohio State University, 191 West Woodruff Avenue, Columbus, OH 43210, USA}

\author[0000-0001-7775-7261]{David~H.~Weinberg}
\affiliation{Department of Astronomy, The Ohio State University, 4055 McPherson Laboratory, 140 W 18th Avenue, Columbus, OH 43210, USA}
\affiliation{Center for Cosmology and AstroParticle Physics, The Ohio State University, 191 West Woodruff Avenue, Columbus, OH 43210, USA}

\author[0000-0003-1887-6732]{Zheng~Zheng}
\affiliation{Department of Physics and Astronomy, The University of Utah, E2108 Stewart Building, 270 S 1400 E, Salt Lake City, UT 84112, USA}


\author{J.~Aguilar}
\affiliation{Lawrence Berkeley National Laboratory, 1 Cyclotron Road, Berkeley, CA 94720, USA}

\author[0000-0001-6098-7247]{S.~Ahlen}
\affiliation{Department of Physics, Boston University, 590 Commonwealth Avenue, Boston, MA 02215 USA}

\author[0000-0001-9712-0006]{D.~Bianchi}
\affiliation{Dipartimento di Fisica ``Aldo Pontremoli'', Universit\`a degli Studi di Milano, Via Celoria 16, I-20133 Milano, Italy}
\affiliation{INAF-Osservatorio Astronomico di Brera, Via Brera 28, 20122 Milano, Italy}

\author{D.~Brooks}
\affiliation{Department of Physics \& Astronomy, University College London, Gower Street, London, WC1E 6BT, UK}

\author{T.~Claybaugh}
\affiliation{Lawrence Berkeley National Laboratory, 1 Cyclotron Road, Berkeley, CA 94720, USA}

\author[0000-0002-2169-0595]{A.~Cuceu}
\affiliation{Lawrence Berkeley National Laboratory, 1 Cyclotron Road, Berkeley, CA 94720, USA}

\author[0000-0002-1769-1640]{A.~de la Macorra}
\affiliation{Instituto de F\'{\i}sica, Universidad Nacional Aut\'{o}noma de M\'{e}xico,  Circuito de la Investigaci\'{o}n Cient\'{\i}fica, Ciudad Universitaria, Cd. de M\'{e}xico  C.~P.~04510,  M\'{e}xico}

\author[0000-0003-4992-7854]{S.~Ferraro}
\affiliation{Lawrence Berkeley National Laboratory, 1 Cyclotron Road, Berkeley, CA 94720, USA}
\affiliation{University of California, Berkeley, 110 Sproul Hall \#5800 Berkeley, CA 94720, USA}

\author[0000-0002-3033-7312]{A.~Font-Ribera}
\affiliation{Institut de F\'{i}sica d’Altes Energies (IFAE), The Barcelona Institute of Science and Technology, Edifici Cn, Campus UAB, 08193, Bellaterra (Barcelona), Spain}

\author[0000-0002-2890-3725]{J.~E.~Forero-Romero}
\affiliation{Departamento de F\'isica, Universidad de los Andes, Cra. 1 No. 18A-10, Edificio Ip, CP 111711, Bogot\'a, Colombia}
\affiliation{Observatorio Astron\'omico, Universidad de los Andes, Cra. 1 No. 18A-10, Edificio H, CP 111711 Bogot\'a, Colombia}

\author[0000-0001-9632-0815]{E.~Gaztañaga}
\affiliation{Institut d'Estudis Espacials de Catalunya (IEEC), c/ Esteve Terradas 1, Edifici RDIT, Campus PMT-UPC, 08860 Castelldefels, Spain}
\affiliation{Institute of Cosmology and Gravitation, University of Portsmouth, Dennis Sciama Building, Portsmouth, PO1 3FX, UK}
\affiliation{Institute of Space Sciences, ICE-CSIC, Campus UAB, Carrer de Can Magrans s/n, 08913 Bellaterra, Barcelona, Spain}

\author{G.~Gutierrez}
\affiliation{Fermi National Accelerator Laboratory, PO Box 500, Batavia, IL 60510, USA}

\author[0000-0001-9822-6793]{J.~Guy}
\affiliation{Lawrence Berkeley National Laboratory, 1 Cyclotron Road, Berkeley, CA 94720, USA}

\author[0000-0002-9136-9609]{H.~K.~Herrera-Alcantar}
\affiliation{Institut d'Astrophysique de Paris. 98 bis boulevard Arago. 75014 Paris, France}
\affiliation{IRFU, CEA, Universit\'{e} Paris-Saclay, F-91191 Gif-sur-Yvette, France}

\author[0000-0002-6550-2023]{K.~Honscheid}
\affiliation{Center for Cosmology and AstroParticle Physics, The Ohio State University, 191 West Woodruff Avenue, Columbus, OH 43210, USA}
\affiliation{Department of Physics, The Ohio State University, 191 West Woodruff Avenue, Columbus, OH 43210, USA}

\author[0000-0002-6024-466X]{M.~Ishak}
\affiliation{Department of Physics, The University of Texas at Dallas, 800 W. Campbell Rd., Richardson, TX 75080, USA}

\author[0000-0003-0201-5241]{R.~Joyce}
\affiliation{NSF NOIRLab, 950 N. Cherry Ave., Tucson, AZ 85719, USA}

\author[0000-0002-8828-5463]{D.~Kirkby}
\affiliation{Department of Physics and Astronomy, University of California, Irvine, 92697, USA}

\author[0000-0003-3510-7134]{T.~Kisner}
\affiliation{Lawrence Berkeley National Laboratory, 1 Cyclotron Road, Berkeley, CA 94720, USA}

\author[0000-0001-6356-7424]{A.~Kremin}
\affiliation{Lawrence Berkeley National Laboratory, 1 Cyclotron Road, Berkeley, CA 94720, USA}

\author[0000-0002-6731-9329]{C.~Lamman}
\affiliation{The Ohio State University, Columbus, 43210 OH, USA}

\author[0000-0003-1838-8528]{M.~Landriau}
\affiliation{Lawrence Berkeley National Laboratory, 1 Cyclotron Road, Berkeley, CA 94720, USA}

\author[0000-0001-7178-8868]{L.~Le~Guillou}
\affiliation{Sorbonne Universit\'{e}, CNRS/IN2P3, Laboratoire de Physique Nucl\'{e}aire et de Hautes Energies (LPNHE), FR-75005 Paris, France}

\author[0000-0002-1125-7384]{A.~Meisner}
\affiliation{NSF NOIRLab, 950 N. Cherry Ave., Tucson, AZ 85719, USA}

\author{R.~Miquel}
\affiliation{Instituci\'{o} Catalana de Recerca i Estudis Avan\c{c}ats, Passeig de Llu\'{\i}s Companys, 23, 08010 Barcelona, Spain}
\affiliation{Institut de F\'{i}sica d’Altes Energies (IFAE), The Barcelona Institute of Science and Technology, Edifici Cn, Campus UAB, 08193, Bellaterra (Barcelona), Spain}

\author{A.~Muñoz-Gutiérrez}
\affiliation{Instituto de F\'{\i}sica, Universidad Nacional Aut\'{o}noma de M\'{e}xico,  Circuito de la Investigaci\'{o}n Cient\'{\i}fica, Ciudad Universitaria, Cd. de M\'{e}xico  C.~P.~04510,  M\'{e}xico}

\author[0000-0003-3188-784X]{N.~Palanque-Delabrouille}
\affiliation{IRFU, CEA, Universit\'{e} Paris-Saclay, F-91191 Gif-sur-Yvette, France}
\affiliation{Lawrence Berkeley National Laboratory, 1 Cyclotron Road, Berkeley, CA 94720, USA}

\author[0000-0002-0644-5727]{W.~J.~Percival}
\affiliation{Department of Physics and Astronomy, University of Waterloo, 200 University Ave W, Waterloo, ON N2L 3G1, Canada}
\affiliation{Perimeter Institute for Theoretical Physics, 31 Caroline St. North, Waterloo, ON N2L 2Y5, Canada}
\affiliation{Waterloo Centre for Astrophysics, University of Waterloo, 200 University Ave W, Waterloo, ON N2L 3G1, Canada}

\author{C.~Poppett}
\affiliation{Lawrence Berkeley National Laboratory, 1 Cyclotron Road, Berkeley, CA 94720, USA}
\affiliation{Space Sciences Laboratory, University of California, Berkeley, 7 Gauss Way, Berkeley, CA  94720, USA}
\affiliation{University of California, Berkeley, 110 Sproul Hall \#5800 Berkeley, CA 94720, USA}

\author[0000-0001-7145-8674]{F.~Prada}
\affiliation{Instituto de Astrof\'{i}sica de Andaluc\'{i}a (CSIC), Glorieta de la Astronom\'{i}a, s/n, E-18008 Granada, Spain}

\author[0000-0001-6979-0125]{I.~P\'erez-R\`afols}
\affiliation{Departament de F\'isica, EEBE, Universitat Polit\`ecnica de Catalunya, c/Eduard Maristany 10, 08930 Barcelona, Spain}

\author{G.~Rossi}
\affiliation{Department of Physics and Astronomy, Sejong University, 209 Neungdong-ro, Gwangjin-gu, Seoul 05006, Republic of Korea}

\author[0000-0002-9646-8198]{E.~Sanchez}
\affiliation{CIEMAT, Avenida Complutense 40, E-28040 Madrid, Spain}

\author{D.~Schlegel}
\affiliation{Lawrence Berkeley National Laboratory, 1 Cyclotron Road, Berkeley, CA 94720, USA}

\author{M.~Schubnell}
\affiliation{Department of Physics, University of Michigan, 450 Church Street, Ann Arbor, MI 48109, USA}
\affiliation{University of Michigan, 500 S. State Street, Ann Arbor, MI 48109, USA}

\author[0000-0002-3461-0320]{J.~Silber}
\affiliation{Lawrence Berkeley National Laboratory, 1 Cyclotron Road, Berkeley, CA 94720, USA}

\author{D.~Sprayberry}
\affiliation{NSF NOIRLab, 950 N. Cherry Ave., Tucson, AZ 85719, USA}

\author[0000-0003-1704-0781]{G.~Tarl\'{e}}
\affiliation{University of Michigan, 500 S. State Street, Ann Arbor, MI 48109, USA}

\author{B.~A.~Weaver}
\affiliation{NSF NOIRLab, 950 N. Cherry Ave., Tucson, AZ 85719, USA}

\author[0000-0002-6684-3997]{H.~Zou}
\affiliation{National Astronomical Observatories, Chinese Academy of Sciences, A20 Datun Road, Chaoyang District, Beijing, 100101, P.~R.~China}








\begin{abstract}
The intergalactic medium (IGM) around quasars is shaped by their dense environments and by their excess ionizing radiation, forming a “quasar proximity zone” whose size and anisotropy depend on the quasar's halo mass, luminosity, age, and radiation geometry. Using over 10,000 quasar pairs from the Dark Energy Spectroscopic Instrument (DESI) Year 1 data, with projected comoving separations $r_{\perp} < 2\,h^{-1}{\rm Mpc}$, we investigate how the proximity zone of foreground quasars at $z\sim2{\rm-}3.5$ affects Lyman-alpha absorption in their background quasars. The large DESI sample enables unprecedented precision in measuring this “transverse proximity” effect, allowing a detailed investigation of the signal's dependence on the projected separation of quasar pairs and the luminosity of the foreground quasar. We find that enhanced gas clustering near quasars dominates over their ionizing effect, leading to stronger absorption on neighboring sightlines. Under the assumption that quasar ionizing luminosity is isotropic and steady, we infer the IGM overdensity profile in the vicinity of quasars, finding overdensities as high as $\Delta \sim 10$ at comoving distance $\sim 1\,h^{-1}{\rm Mpc}$ from the most luminous systems. Surprisingly, however, we find no significant dependence of the proximity profile on the luminosity of the foreground quasar. This lack of luminosity dependence could reflect a cancellation between higher ionizing flux and higher gas overdensity, or it could indicate that quasar emission is highly time variable or anisotropic, so that the observed luminosity does not trace the ionizing flux on nearby sightlines.


\end{abstract}



\section{Introduction} \label{sec:intro}

Active galactic nuclei (AGN) play a crucial role in understanding the co-evolution of galaxies and their central supermassive black holes \citep[e.g.,][]{2013ARA&A..51..511K}. In particular, quasars — an extremely luminous phase of AGN — emit intense radiation that can significantly influence both the intergalactic medium (IGM) and nearby galaxies \citep[e.g.,][]{2009RvMP...81.1405M}.
This connection allows us to use the large-scale ($\gtrsim 1\Mpc$)  impact of quasar radiation as a probe of the compact central regions ($\lesssim 10\pc$) that host the AGN and their central supermassive black holes. Quasars thus serve as a powerful observational tool to investigate the interplay between black hole activity and galaxy-scale environments across cosmic time.

Among the observational signatures of quasars, two aspects are particularly informative for probing the small-scale physics of AGN: the {\it timescale} of their emission and its {\it spatial anisotropy}.
First, quasar lifetimes remain highly uncertain, with current population-averaged estimates ranging from $10^{5}$ to $10^{8}\mathrm{yr}$ \citep[e.g.,][]{2021ApJ...917...38E}. In contrast, the recent discovery of changing-look quasars — objects that exhibit dramatic spectral transitions over just a few years — suggests that at least some quasars may vary an enormous amount on surprisingly short timescales \citep[e.g.,][]{2016MNRAS.457..389M}. Understanding the relationship between these two vastly different timescales offers a unique window into the physical conditions of the accretion disk and its immediate surroundings.
Second, the AGN unification model predicts that quasar radiation is anisotropic, due to a dusty torus that obscures the central engine depending on the viewing angle \citep[e.g.,][]{2015ARA&A..53..365N}. This anisotropy has been confirmed through comparisons of \lya absorption in the longitudinal and transverse directions relative to quasars \citep[e.g.,][]{2013ApJ...776..136P}. The geometry and orientation of this obscuring structure determine whether a quasar appears as type 1 (unobscured) or type 2 (obscured), thereby providing direct insight into the nuclear structure on parsec scales.

The impact of quasar ionizing radiation on the surrounding IGM, known as the quasar proximity effect \citep{1988ApJ...327..570B}, has been widely used to probe quasar activity in a model-independent way \citep[see, e.g.,][for a review]{2019ApJ...882..165S}. Since the size of the ionized region depends on the quasar’s age and the propagation speed of the ionization front, proximity effect measurements can constrain quasar lifetimes without relying on assumptions about their demographics (e.g., luminosity function or host halo mass). Two forms of this effect have been studied: the line-of-sight proximity effect (LPE), observed as a reduction in \lya forest absorption in the quasar’s own spectrum \citep{1982MNRAS.198...91C,1988ApJ...327..570B}, and the transverse proximity effect (TPE), identified by \lya absorption features along background sightlines passing near foreground quasars \citep{1989ApJ...336..550C,1991ApJ...377L..69D,2004ApJ...612..706A}. 

Recent studies of the quasar proximity effect have focused on the LPE at high redshifts ($z \gtrsim 5$), in the context of cosmic reionization or the formation of supermassive black holes \citep[for a recent summary, see][]{2025arXiv250509676O}. At these redshifts, the LPE is often discussed under the name “near zone,” here taken to include cases in which the observed high-transmission region blueward of \lya is set by a quasar's \ion{H}{2} ionization front propagating into a substantially neutral IGM \citep[e.g.,][]{2006AJ....132..117F,2007MNRAS.374..493B}. One key reason for this focus is that the high neutral hydrogen fraction at that epoch makes it easier to detect the ionizing impact of quasars on the surrounding IGM. In addition, for such high-redshift quasars, obtaining background spectra at small angular separations is observationally challenging, making the LPE a more accessible approach. Nonetheless, a recent JWST study has demonstrated the feasibility of detecting the TPE even at $z\sim6.3$, where \lya tomography of 12 background galaxies revealed a quasar “light echo” and provided new constraints on the geometry of its ionization cone and on the timescale of the quasar’s UV radiation \citep{2025arXiv250905417E}.

At redshifts $2 \lesssim z \lesssim 4$, the TPE becomes more accessible, as background spectra are more readily available. Moreover, while the LPE provides only lower limits on quasar age in this redshift range due to the short ionization timescale, the TPE offers a complementary advantage: it enables us to probe much longer timescales, since it is sensitive to radiation emitted in the past \citep[for theoretical context, see][]{2024MNRAS.531.2912H}. This redshift range encompasses the cosmic peak of quasar activity. Understanding quasar physics during this active phase provides essential clues to the co-evolution of galaxies and black holes, including the connection between galaxy morphology, star formation rates, and AGN feedback processes.

Several studies have investigated the TPE at moderate redshifts ($2 \lesssim z \lesssim 4$) using quasar pair samples to probe quasar radiation properties. \citet{2004ApJ...610..642C} predicted decreased \lya absorption near foreground quasars using simulations, but instead found enhanced absorption, suggesting short episodic lifetimes or anisotropic emission. The Quasars Probing Quasars (QPQ) survey similarly reported excess neutral hydrogen (\ion{H}{1}) absorption and a high covering fraction of optically thick systems around $z \sim 2$ quasars \citep[e.g.,][]{2013ApJ...776..136P}, consistent with dense environments and obscured radiation. In contrast, \citet{2008MNRAS.391.1457K} found no TPE signal in over 100 quasar pairs, inferring lifetimes shorter than $\sim 1\ {\rm Myr}$.

TPE has also been explored using \ion{He}{2} \lya absorption at similar redshifts \citep[e.g.,][]{2017ApJ...847...81S,2018ApJ...861..122S}. Due to the much higher \lya opacity of \ion{He}{2} compared to \ion{H}{1} in this redshift range, variations in absorption can be more dramatic, analogous to the \ion{H}{1} \lya forest at higher redshifts. However, these studies face challenges such as the limited number of available \ion{He}{2} sightlines, which number only a few dozen \citep[e.g.,][]{2019ApJ...882..165S}, and the inhomogeneous ionization state of \ion{He}{2} in the IGM around $z \sim 3$, which complicate the interpretation.
While these results highlight the potential of TPE as a probe of quasar variability and geometry, the observational outcomes remain inconclusive, reflecting a complex interplay between episodic activity, anisotropic emission, and the properties of the surrounding environment, including local gas density.

The influence of quasar ionizing radiation on the local intergalactic environment has been studied in large-scale structure (LSS) analyses, where cross-correlations between quasars and the \lya forest aim to probe the underlying matter density distribution. This radiative contribution was thoroughly explored in \citet{2013JCAP...05..018F}, and non-zero signals have since been measured in spectroscopic surveys \citep[e.g.,][]{2020ApJ...901..153D}. In particular, recent analyses of the Dark Energy Spectroscopic Instrument \citep[DESI;][]{2016arXiv161100036D,2016arXiv161100037D} data have reported non-zero TPE amplitudes under the simplifying assumption of isotropic quasar emission, which makes the modeling more tractable. In these analyses, the effect is captured by a single free parameter, $\xi^{\rm TP}_0$, which scales the additional term introduced to represent the TPE. Reported values are $\xi^{\rm TP}_0 = 0.395 \pm 0.051$ and $0.453 \pm 0.046$ for DR1 \citep{2025JCAP...01..124A} and DR2 \citep{2025arXiv250314739D}, respectively. These results underscore the importance of accurately modeling both quasar radiation and their local environments when interpreting \lya forest clustering measurements.

Independent efforts using the LPE have also aimed to characterize the dense environments around quasars. Several studies have analyzed the distribution of \lya optical depth in proximity zones and shown that the observed reduction in absorption is weaker than expected from ionizing radiation alone, which suggests the presence of significant overdensities. For example, \citet{2005MNRAS.361.1015R} and \citet{2007MNRAS.377..657G} applied optical depth statistics to high-resolution quasar spectra at $z \sim 2$–$4$ and found mean overdensities of a factor of $\sim$2–5 within $\sim 10\hinvmpc$. These results support the picture that quasars reside in highly biased regions likely to evolve into massive galaxy clusters. Furthermore, \citet{2008ApJ...673...39F} demonstrated that the LPE signal in cosmological simulations depends on the mass of quasar host halos due to both local overdensities and gas infall. They showed that neglecting these effects can bias the inferred ionizing background by up to a factor of $\sim$2.5 at $z=3$, and they proposed that proximity effect measurements can be inverted to estimate quasar host halo masses.

A recent study by \citet{2019ApJ...884..151J} utilized a large sample of 181 projected quasar pairs from the Sloan Digital Sky Survey (SDSS) DR12 to investigate the quasar proximity effect in both the longitudinal and transverse directions. Their work demonstrated the power of using quasar pairs from wide-area spectroscopic surveys to statistically probe quasar environments and radiation geometry. In this study, we focus on the TPE using the DESI Year 1 (Y1) data publicly available as part of the first data release \citep[DR1,][]{2025arXiv250314745D}, which provides an unprecedentedly large sample of quasar pairs. Specifically, the DESI DR1 catalog includes 531,000 quasars at $z>2.1$ for \lya forest studies, nearly triple the 184,101 quasars with $z>2.15$ available in SDSS DR12 \citep{2017A&A...597A..79P}. Leveraging this sample size, we investigate the redshift, luminosity, and transverse-distance dependence of the TPE by constructing sub-samples, allowing us to examine the physical nature of quasar radiation and its impact on the surrounding IGM in greater detail. Our full sample includes more than 10,000 quasar pairs with transverse separations $r_{\perp} < 2 \hinvmpc$ (comoving).

The structure of this paper is as follows. In Section~\ref{sec:data}, we describe the data used in this study, including the DESI Y1 quasar sample, the extraction of the \lya forest absorption spectra, and the construction of quasar pair and control samples. Section~\ref{sec:results} presents our measurements of the stacked \lya transmission near quasars, based on three types of subsamples constructed by redshift, luminosity, and transverse separation. In Section~\ref{sec:interpretation}, we interpret the observed signals by modeling the \ion{H}{1} optical depth in the proximity region, estimate the impact of quasar radiation and gas overdensity, and test for anisotropy in quasar emission. Finally, we summarize our findings in Section~\ref{sec:summary}.
Throughout this paper, we assume a flat $\Lambda$CDM cosmology consistent with the best-fit parameters from the Planck 2018 results \citep[][]{2020A&A...641A...6P}, with $H_0 = 67.32\kmsMpc$ and $\Omega_{\rm m} = 0.3158$. We use comoving distances unless otherwise noted.

\section{Data} \label{sec:data}

We use quasar spectra obtained during the first year of the DESI main survey, which ran from May 14, 2021, to June 14, 2022. These data were obtained with extensive new instrumentation that was designed to observe 5000 spectra in a single observation \citep{2022AJ....164..207D}. The instrumentation includes a new, 3 square degree prime focus corrector \citep{2024AJ....168...95M}, a focal-plane system with robotic fiber positioners \citep{2023AJ....165....9S}, and 5000 fiber optics cables that direct light \citep{2024AJ....168..245P} into ten, bench-mounted spectrographs. Each of the ten spectrographs has three channels that together span from 3600\AA\ to 9800\AA\ and have a spectral resolution that ranges from 2000 to 5000. The DESI spectroscopic pipeline is described in \citet{2023AJ....165..144G} and the nightly survey operations is described in \citet{2023AJ....166..259S}. The processed products were released as the DESI Year 1 data set in the first public data release \citep[DESI DR1;][]{2025arXiv250314745D}, forming the foundation for subsequent scientific analyses, including the first cosmological constraints from clustering measurements with DESI \citep{2025JCAP...07..028A}.

\subsection{Quasar Catalog}

The target selection, validation, and classification of DESI quasars are described in \citet{2023ApJ...944..107C}. In particular, quasar classification and redshift determination follow a multi-step procedure to ensure high completeness and accuracy. The primary tool is the template-fitting code {\tt Redrock}\footnote{\url{https://github.com/desihub/redrock/}} \citep{Redrock.Bailey.2025}, which fits PCA-based templates for stars, galaxies, and quasars across a wide redshift range and assigns the most likely class and redshift by minimizing $\chi^{2}$. To further improve completeness and redshift accuracy, two “afterburners” were introduced \citep[for details, see][]{2023AJ....165..124A}: one that searches for broad \ion{Mg}{2} emission in spectra not initially classified as quasars, and {\tt QuasarNET}\footnote{\url{https://github.com/ngbusca/QuasarNET}} \citep{2018arXiv180809955B,2020JCAP...11..015F}, a deep convolutional neural network trained to identify quasar emission features and provide redshift estimates. For the Ly$\alpha$ forest analysis, we further adopt updated redshifts based on improved high-$z$ quasar templates, which mitigate biases identified in the original catalog \citep{2023AJ....166...66B,2025JCAP...01..130B}. For quasar luminosities, we utilize the monochromatic luminosity at 1500\AA\ in the rest frame measured by {\tt fastspecfit}\footnote{\url{https://github.com/desihub/fastspecfit}} \citep{2023ascl.soft08005M}.

\subsection{\lya forest}

The \lya forest catalog, a subset of the quasar spectra, is described in \citet{2024MNRAS.528.6666R} and \citet{2025JCAP...01..124A}. Here, we summarize its main characteristics and properties. There are two different catalogs that correspond to two different rest-frame wavelength ranges of the quasar spectra, Region “A” (1040--1205\AA) and Region “B” (920--1020\AA). In the following analysis, we focus on the Region A catalog, which is not affected by other Lyman series lines. Note that we use only data from the blue channel (one of the three spectrograph channels), and the observed wavelength coverage is limited to 3600--5772\AA. Combined with the above rest-frame wavelength cut, this corresponds to a quasar redshift range of $z=2.1$--$4.4$ for Region A. 

Furthermore, various masks are applied to remove bad pixels caused by cosmic rays or CCD defects, as well as to mitigate galactic absorption, residual sky lines, and the effects of damped \lya systems (DLAs) and broad absorption line (BAL) features. Briefly, DLAs are identified using both convolutional neural network (CNN) and Gaussian Process (GP)-based finders, and regions around DLAs detected by both algorithms with probability $>50\%$ are masked. 
To ensure purity, the DLA catalog only includes DLAs identified in spectra with mean Ly$\alpha$-forest S/N $>3$. In our selection of \lya forests or quasar pairs, we do not impose such an S/N cut. Since the DLA-masked pixels are only a small fraction \citep[$\sim 5\%$; e.g., Figure~3 of][]{2024MNRAS.528.6666R} in the DESI \lya forest catalogs, their overall impact on the measured transmission is small.
BAL features are identified using the DESI DR1 BAL catalog, which is based on the Absorptivity Index (AI) and Balnicity Index (BI) criteria; 
the corresponding absorption troughs are masked, while the rest of the \lya forest region in those spectra is retained. After discarding forests that do not have 150 valid pixels (too short) or do not have a valid continuum fit (negative continuum), we finally have 428,403 \lya forests (Region A).

The catalog provides the density fluctuation and the associated weight for each forest pixel. The density fluctuation is measured by the DESI \lya analysis pipeline, {\sc picca}\footnote{\url{https://github.com/igmhub/picca/}} \citep{2020ApJ...901..153D}, and defined by the transmitted flux:
\begin{eqnarray}
    \delta_{q}(\lambda) = \frac{f_{q}(\lambda)}{\overline{F}(\lambda)C_{q}(\lambda)} -1, \label{eq:delta_q}
\end{eqnarray}
where $f_{q}(\lambda)$ is the observed flux, $\overline{F}(\lambda)$ is the mean transmission, and $C_{q}(\lambda)$ is the unabsorbed continuum of the quasar spectrum. The subscript $q$ references an individual quasar, while $\overline{F}$ is an average over all quasars. We estimate the mean expected flux $\overline{F}(\lambda)C_{q}(\lambda)$ by a continuum fitting procedure, described in detail in \citet{2024MNRAS.528.6666R}. The main assumption behind this procedure is that $\overline{F}(\lambda)C_{q}(\lambda)$ for each forest can be expressed as a universal function of the quasar continuum, $\overline{C}(\lambda_{\rm rf})$, adjusted by a linear polynomial in $\Lambda \equiv \log \lambda$. In addition to its fitting parameters, the total variance of $f_{q}$, $\sigma_{q}^{2}$, is determined through likelihood maximization in the following form:
\begin{eqnarray}
    \sigma_{q}^{2}(\lambda) = \eta_{\rm pip}(\lambda)\sigma^{2}_{{\rm pip}, q}(\lambda) + \sigma_{\rm LSS}^{2}(\lambda)(\overline{F}C_{q})^{2}(\lambda),  \label{eq:sigma_q}
\end{eqnarray}
where the first term represents the pipeline noise, $\sigma^{2}_{{\rm pip}, q}$, modulated by a function $\eta_{\rm pip}$, while the second term corresponds to the intrinsic variance of the \lya forest, given by $(\overline{F}C_{q})^{2}$ scaled by a universal function $\sigma_{\rm LSS}^{2}$. The weight associated with the transmitted flux (Equation~\ref{eq:delta_q}) is then given as the inverse of the variance,
\begin{eqnarray}
    w_{\rm q}(\lambda) = \frac{1}{\eta_{\rm pip}(\lambda)\tilde{\sigma}^{2}_{{\rm pip}, q}(\lambda) + \eta_{\rm LSS}\sigma_{\rm LSS}^{2}(\lambda)},  \label{eq:weight_q}
\end{eqnarray}
where $\tilde{\sigma}_{{\rm pip}, q} = \sigma_{{\rm pip}, q}/\overline{F}C_{q}$. An additional scaling parameter, $\eta_{\rm LSS}$, was introduced to optimize the correlation function measurement \citep{2024MNRAS.528.6666R}, modulating the contribution of the intrinsic variance, and was fixed at a value of 7.5 for this catalog.\footnote{The value of $\eta_{\rm LSS} = 7.5$ was determined to optimize the precision of the auto- and cross-correlation of the density fluctuation ($\delta_q$, Equation~\ref{eq:delta_q}), rather than that of the transmitted flux ($f_q / C_q$), which is the focus of this paper. Nevertheless, we expect this choice to have no significant impact on our qualitative error estimation, as the gain in precision for cross-correlation is only 10\% compared to the case with $\eta_{\rm LSS} = 1.0$ \citep{2024MNRAS.528.6666R}.} Since the above expression for the weight corresponds to $(\overline{F}C_{q})^{2}/\sigma_{q}^{2}$, apart from the factor $\eta_{\rm LSS}$, we evaluate the S/N for each pixel by relating its weight to the squared S/N as $w_{\rm q}(\lambda)=({\rm S/N})^{2}(\lambda)$.

\subsection{Quasar pairs} \label{sec:qso_pairs}

\begin{figure*}[ht!]
\centering
    \includegraphics[width=\linewidth]{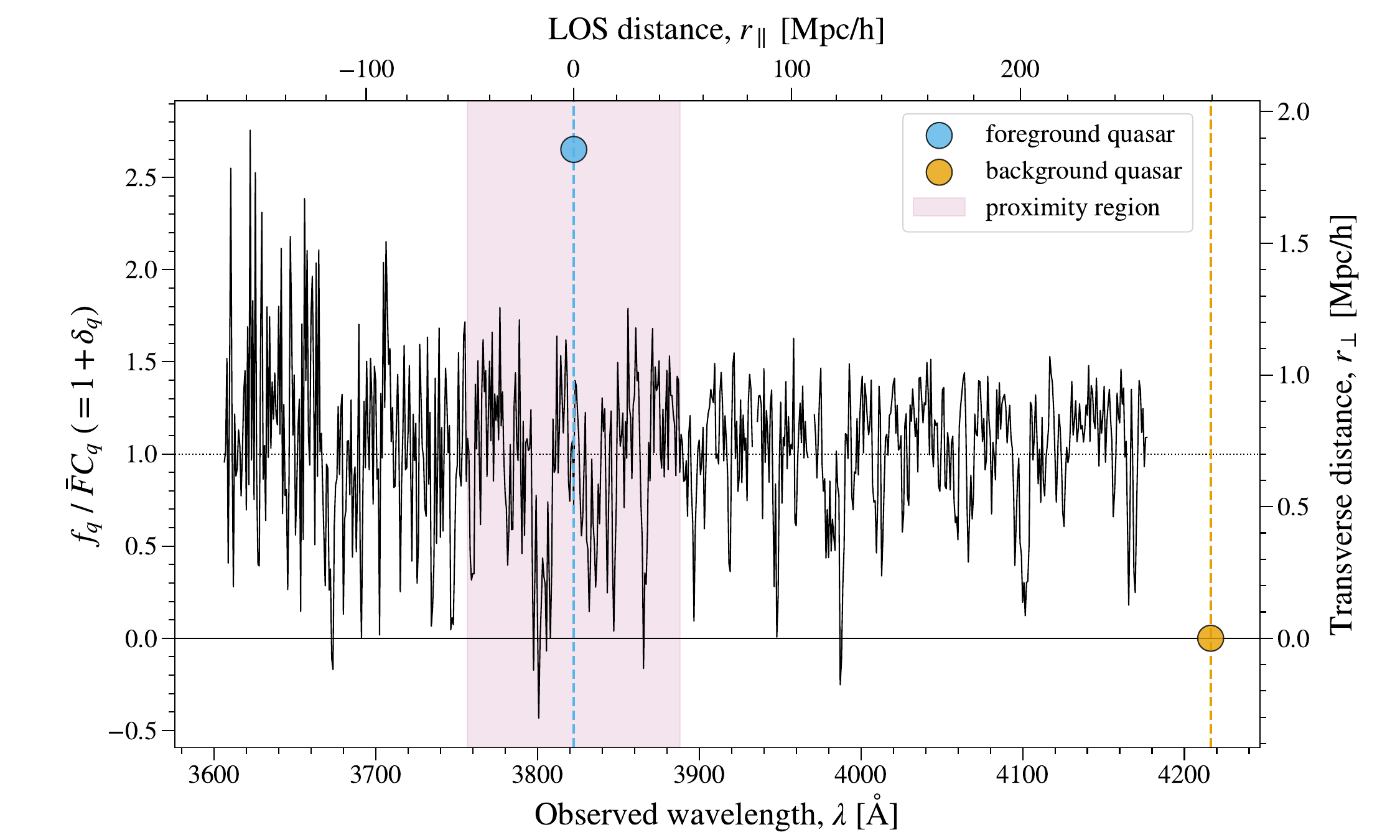} 
\caption{Spectrum of the \lya forest of a background quasar that has a minimum projected separation of $1.86 \hinvmpc$ from a foreground quasar. The sky blue and orange circles indicate the spatial positions of the foreground and background quasars, respectively. The dashed vertical lines mark the \lya wavelengths of these quasars. The pink-shaded region represents the proximity region, defined as a comoving line-of-sight separation of $\pm 50\hinvmpc$ relative to the foreground quasar. The foreground quasar is at $z=2.14$ (TargetID $= 39627480138519001$) and the background quasar is at $z=2.47$ (TargetID $= 39627480138519447$).
\label{fig:qso_pair_spectrum}}
\end{figure*}

Since we are interested in \lya absorption in the vicinity of quasars, we first construct a catalog of closely separated quasar pairs on the sky from the \lya forest catalog. The steps are as follows:

\begin{enumerate}
  \item Consider quasar pairs, consisting of a foreground and a background quasar, selected from a total of 428,403 quasars that constitute the \lya forest catalog, and apply a cut based on the transverse distance between the foreground quasar (at redshift $z_{\rm f/g}$) and the line of sight to the background quasar, requiring \begin{eqnarray}
    D_{\rm M}(z_{\rm f/g})\sin\theta_{\rm sep}<2\hinvmpc,
  \end{eqnarray}
  where $D_{\rm M}(z)$ is the angular (transverse) comoving distance at a redshift $z$ and $\theta_{\rm sep}$ is the separation angle between the quasar pairs. We restrict our analysis to transverse comoving distances $< 2\hinvmpc$ because, for typical quasar luminosities in our sample, the photoionization rate from the foreground quasar falls below the metagalactic UV background (UVB) at larger separations, making the TPE undetectable beyond this scale (see Section~\ref{sec:qso_ionization} for details).
  \\ --- remaining pairs: 25,210
  
  \item Require that the quasar proximity region, within $\pm 50\hinvmpc$ of foreground quasar along the line of sight,\footnote{Although the term {\it quasar proximity region} generally refers to the region around a quasar where its ionizing radiation dominates over the UV background, here we adopt a fixed $\pm 50\hinvmpc$ window purely for convenience in stacking. This choice also encompasses the broad \ion{Si}{3} $1206.5\text{\AA}$ absorption feature (around $-30 \lesssim r_\parallel \lesssim -10\hinvmpc$), clearly seen for the closest transverse sightlines (Figure~\ref{fig:sub_all}).} is fully covered by both the observed wavelength (i.e., 3600--5772\AA) and the \lya forest region in the background quasar spectrum (i.e., the rest-frame wavelength range of 1040--1205\AA).
  \\ --- remaining pairs: 10,942
  
  \item Remove pairs in which the foreground quasar does not have a well-measured luminosity.
  \\ --- remaining pairs: 10,919
  
  \item For pixels in background quasar spectrum within the proximity region of its foreground quasar, consider the fraction of well-defined pixels (e.g., not masked) and the mean S/N averaged over those pixels, given by $\langle({\rm S/N})^2\rangle_{\rm prox}=\langle w_{\rm q}\rangle_{\rm prox}$, and then remove pairs with more than half of the pixels being ill-defined and/or extremely low S/N, $\langle({\rm S/N})^2\rangle_{\rm prox}<0.02$.
  \\ --- remaining pairs: 10,263
\end{enumerate}

Figure~\ref{fig:qso_pair_spectrum} presents an example of the selected quasar pairs, illustrating the spatial positions of the foreground and background quasars, as well as the normalized \lya flux, $f_{q}/\overline{F}C_{q}(=1+\delta_{q})$, of the background quasar spectrum passing through the proximity region of the foreground quasar. Notably, the number of quasar pairs in our sample is nearly 50 times larger than in the \citet{2019ApJ...884..151J} analysis based on the SDSS DR12 quasar catalog. While the number of quasars available for \lya forest analyses increased by only a factor of $\sim$3, this would translate into an expected $\sim$10-fold increase in pair counts, still far below the factor of 50 actually achieved. This additional enhancement arises mainly because they also investigated the LPE and therefore conservatively restricted their quasars to $z>2.57$ in order to ensure spectral coverage of the \lya absorption region.

\subsection{Control sample} \label{sec:control_sample}

The \lya forests observed toward the background quasars in our 10,263 close pairs are expected to be influenced by radiation from their foreground quasars. To investigate variations in quasar proximity regions, we require a reference sample catalog for comparison with the background quasar spectra. Taking advantage of the large DESI dataset, we construct the control sample catalog from the same \lya forest catalog, partially following the procedure adopted in \citet{2019ApJ...884..151J},\footnote{They adopted a smaller proximity region ($\sim40\hinvmpc$ at $z\sim3$) and a relaxed S/N matching criterion of 0.15 in Equation~\ref{eq:cs_StoN}, requiring 25 control sample quasars per foreground quasar.} in which the control sample was selected to match the redshift and S/N of the background quasar spectra.

First, we divide the 428,403 quasars in the \lya forest catalog into 50 redshift bins, each containing an equal number of quasars. For each quasar pair, we then randomly select 10 quasars from the same redshift bin as the background quasar without repetition, ensuring the following condition:
\begin{eqnarray}
    \left|\frac{[{\rm S/N}]_{\rm cs}-[{\rm S/N}]_{\rm b/g}}{[{\rm S/N}]_{\rm b/g}}\right| < 0.1, \label{eq:cs_StoN}
\end{eqnarray} 
where $[{\rm S/N}]\equiv \sqrt{\langle({\rm S/N})^2\rangle_{\rm prox}}=\sqrt{\langle w_{\rm q}\rangle_{\rm prox}}$ is the mean S/N averaged over pixels within the proximity region of the foreground quasar. The subscripts “b/g” and “cs” refer to the background quasar and a candidate quasar to be selected as a control sample, respectively. Note that when computing $[{\rm S/N}]_{\rm cs}$, we exclude quasar spectra in which all pixels within the proximity region are ill-defined.

\section{Results} \label{sec:results}

In this section, we first introduce three types of subsample catalogs and then present the measurements of \lya transmission in the quasar proximity for each. Before presenting the results, we define the line-of-sight and transverse distance of pixels in background quasars at $z_{\rm pix}$ relative to the positions of their foreground quasars: 
\begin{eqnarray}
    r_{\parallel} &=& D_{\rm C}(z_{\rm pix})-D_{\rm C}(z_{\rm f/g})\cos\theta_{\rm sep}
    \nonumber \\
    r_{\perp} &=& D_{\rm M }(z_{\rm f/g})\sin\theta_{\rm sep},
\end{eqnarray}
where $D_{\rm C}$ and $D_{\rm M}$ are the comoving distance and the angular (transverse) comoving distance, respectively, and $\theta_{\rm sep}$ is the separation angle between the quasar pairs. Note that, by definition, all pixels in the same background quasar spectrum share the same transverse distance.\footnote{The conventional definition of distance used in \lya forest auto- and cross-correlation analyses takes a different form \citep[e.g., see Equation~3.1 of][]{2025JCAP...01..124A}, as it focuses on the correlation between pixels (or quasar positions) rather than between {\it sightlines} and quasar positions, which is more suitable for our context here.}

\subsection{Subsamples} \label{sec:subsamples}

\begin{figure*}[ht!]
\centering

\gridline{
  \fig{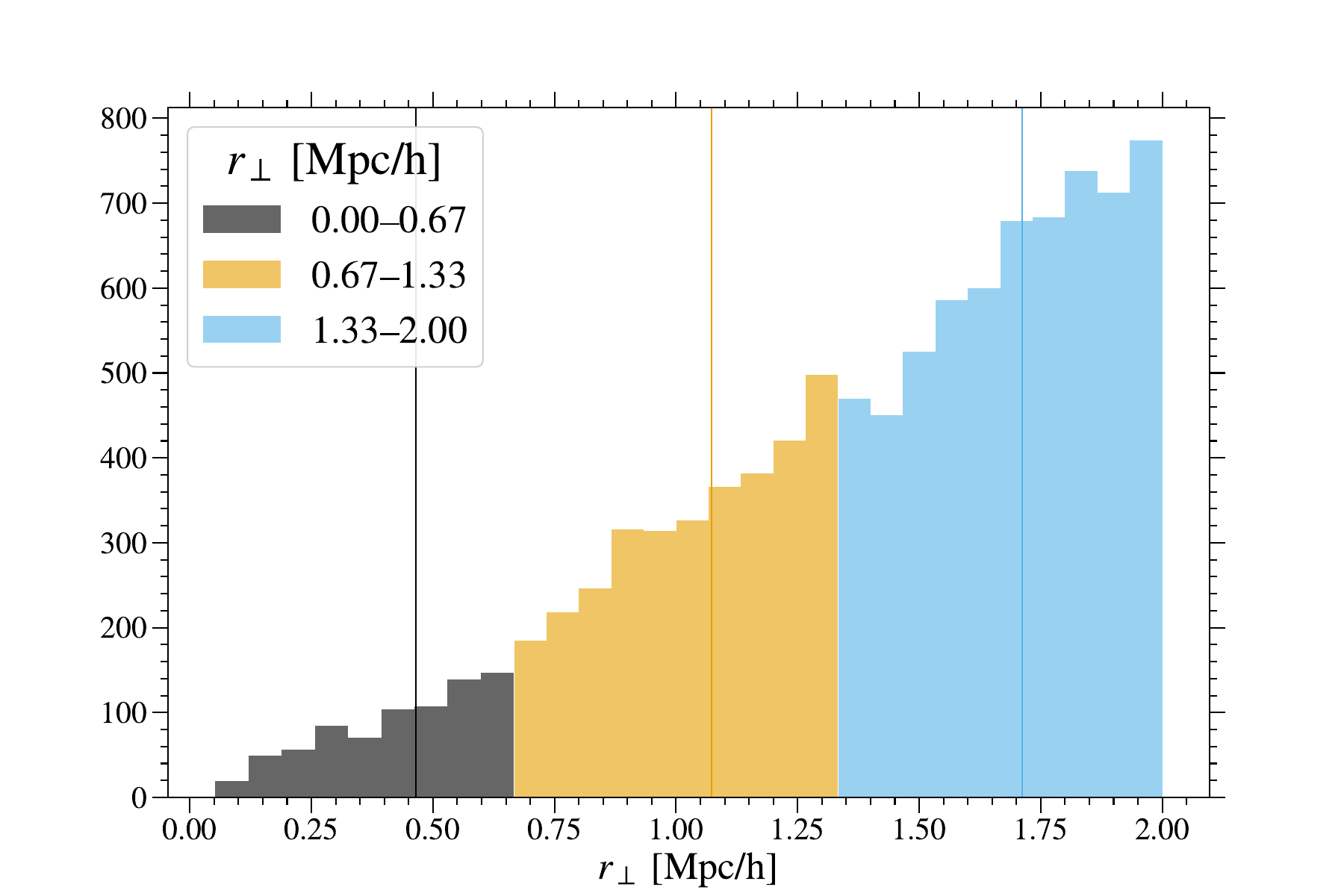}{0.5\textwidth}{(a1)} 
  \fig{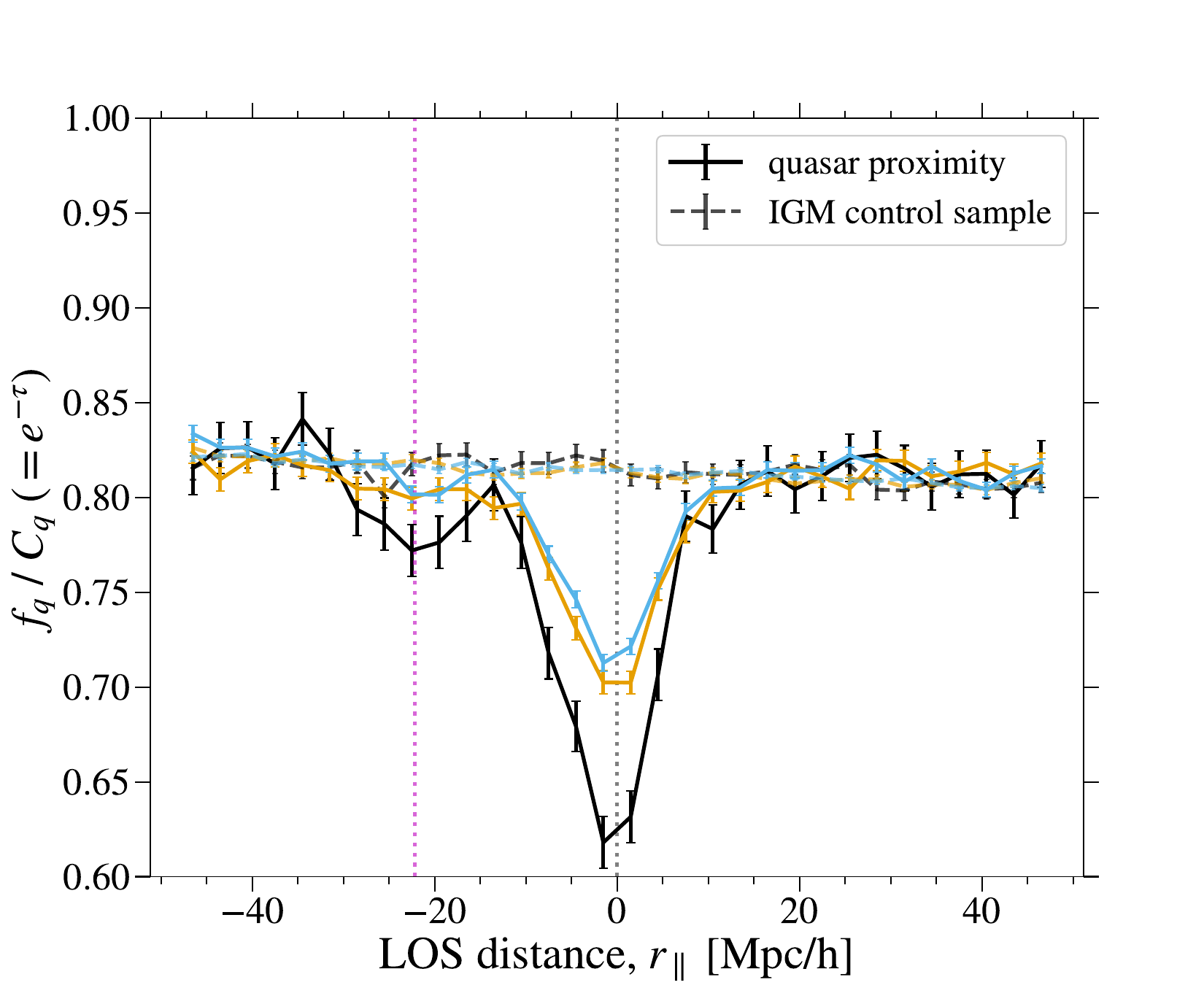}{0.45\textwidth}{(a2)}
}
\vspace{-1.5em}

\gridline{
  \fig{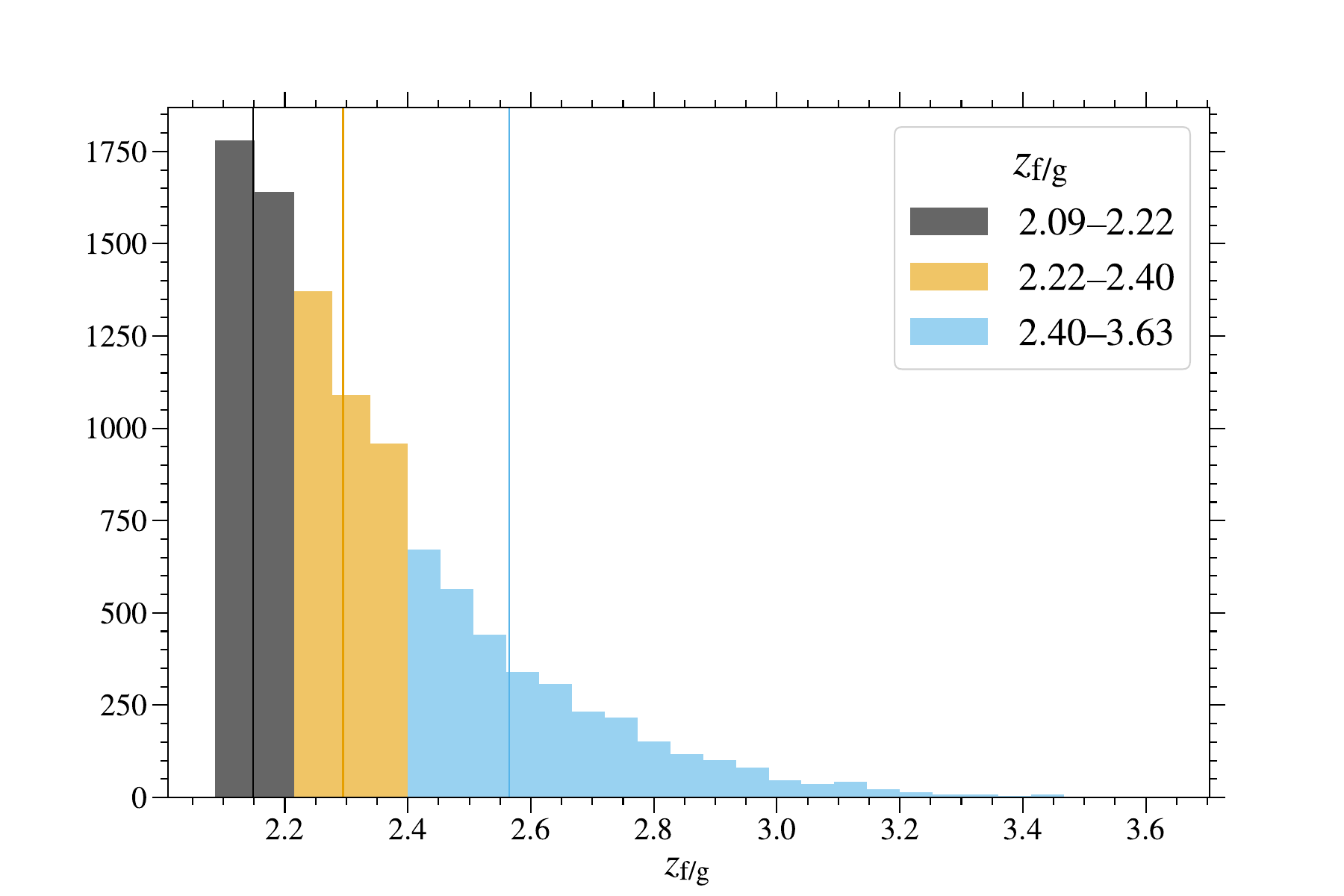}{0.5\textwidth}{(b1)}
  \fig{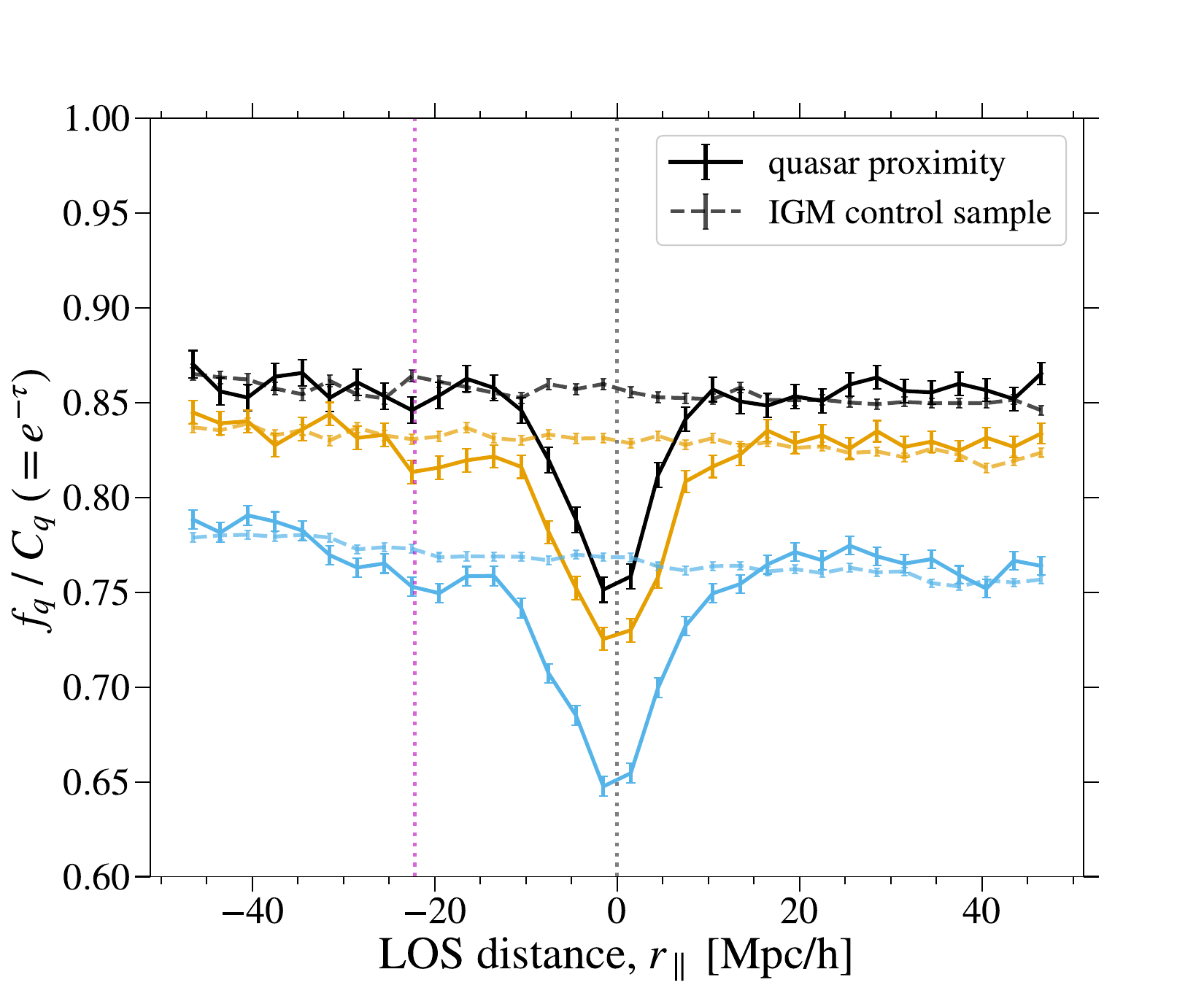}{0.45\textwidth}{(b2)}
}
\vspace{-1.5em}

\gridline{
  \fig{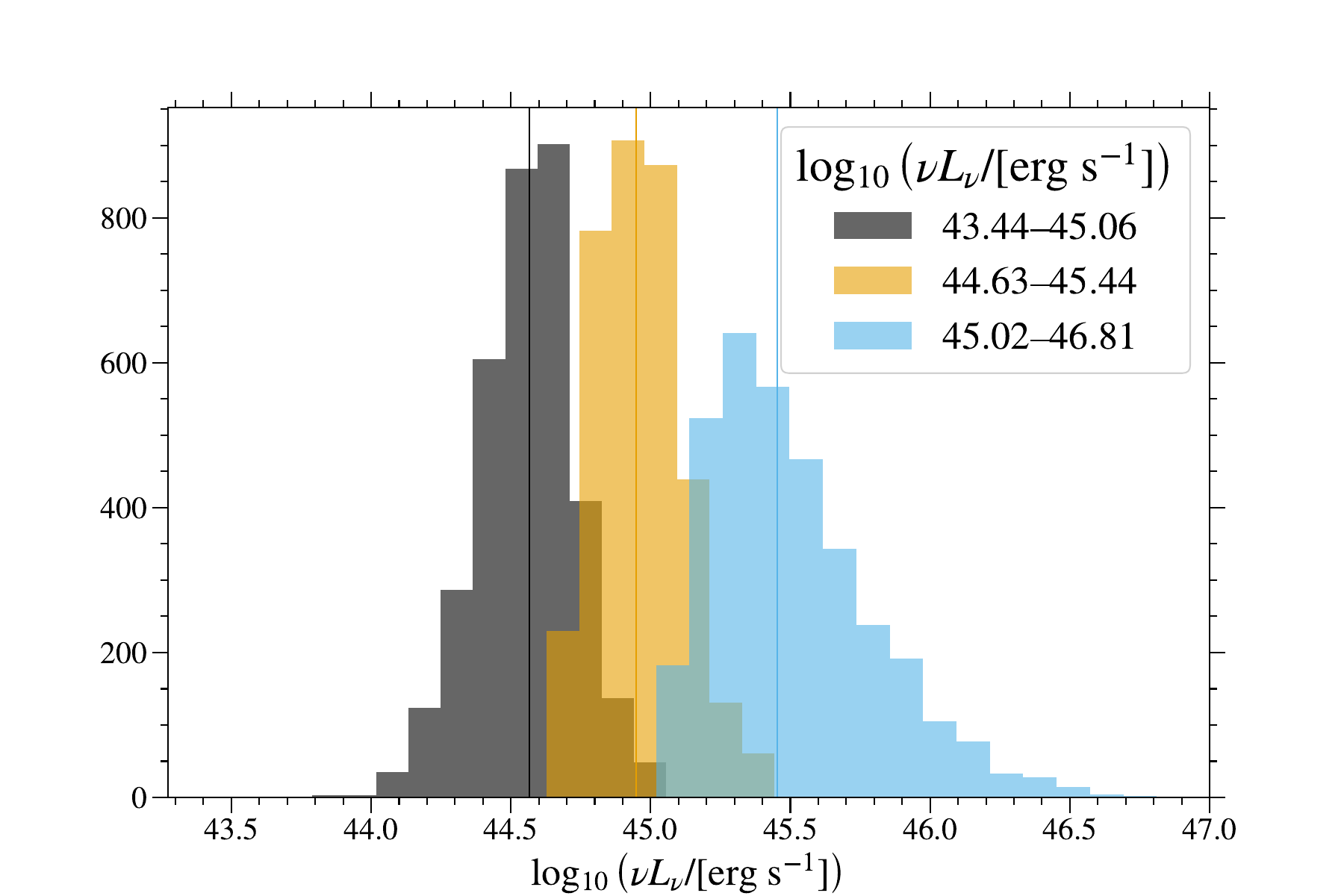}{0.5\textwidth}{(c1)}
  \fig{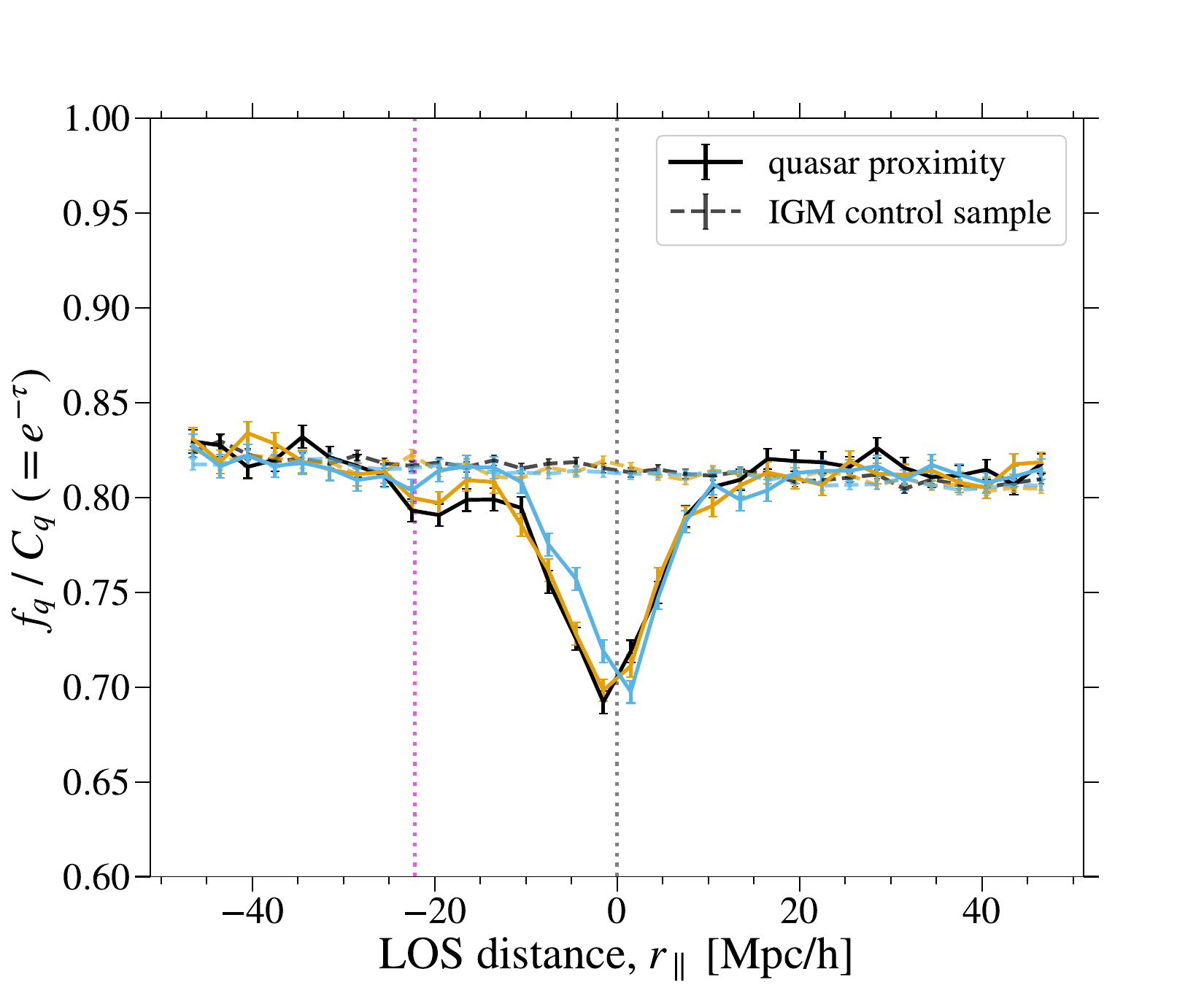}{0.45\textwidth}{(c2)}
}

\caption{Distributions and stacked \lya transmissions for three subsamples divided by transverse distance (top), redshift (middle), and luminosity (bottom).
{\it Left}: Histograms of each subsample, with solid vertical lines showing the median values. 
{\it Right}: Stacked \lya transmissions around foreground quasars (solid) and corresponding control samples (dashed) as a function of line-of-sight separation from the foreground quasars (black dotted line). The pink dotted line marks the \ion{Si}{3} absorption at $1206.5\text{\AA}$. Redshift and luminosity subsamples include all transverse separations $r_\perp = 0$–$2\hinvmpc$, and luminosity bins are selected to have similar redshift distributions. \label{fig:sub_all}
}
\end{figure*}

In the following analysis, we consider three different subsample catalogs based on the transverse distance of the background quasar sightline, the foreground quasar redshift, and the foreground quasar luminosity. The transverse distance of a background quasar sightline directly reflects its susceptibility to the influence of its foreground quasar environment. We divide the quasar pairs into three bins of equal width based on the transverse distance, covering the range 0--2$\hinvmpc$. Panel (a1) of Figure~\ref{fig:sub_all} shows the subsample distribution and illustrates the number of samples scaling with the transverse distance, as expected. We also consider a subsample binned by the foreground quasar redshifts to investigate the redshift evolution of the proximity region. Panel (b1) of Figure~\ref{fig:sub_all} shows the distribution of this subsample, which is divided into three bins containing an equal number of quasars. One of the main interests here is to understand the relationship between variations of \lya absorption in the foreground quasar proximity region and their ionizing radiation flux. Panel (c1) of Figure~\ref{fig:sub_all} represents three bins based on the monochromatic luminosity at rest-frame 1500\AA\ of foreground quasars, $\nu L_\nu |_{1500\text{\AA}}$. Since luminosity and redshift are generally correlated, we divide the sample in a way that preserves similar redshift distributions across luminosity subsamples. Specifically, we first sort the foreground quasars by redshift and divide them into 30 redshift bins. Within each bin, we then divide the quasars into three equal-sized subsamples based on their luminosity. This procedure ensures that the redshift distribution is nearly identical across the luminosity subsamples, although it results in some overlap in luminosity between the bins. This allows us to isolate the dependence on luminosity more clearly without contamination from redshift evolution effects. We summarize in Table~\ref{tab:ranges} the parameter ranges used to define the subsamples based on transverse distance, redshift, and luminosity. We have verified that, as expected, the transverse-distance subsamples have similar distributions in luminosity and redshift, the redshift subsamples have comparable distributions in transverse distance, and the luminosity subsamples also show similar distributions in transverse distance. These results indicate that, apart from the known correlation between redshift and luminosity, no significant correlations are present among these quantities, and they are not expected to bias the comparison among the subsamples.

{\renewcommand{\arraystretch}{1.5}
\begin{table}[t!]
\caption{Parameter ranges for subsamples of 10,263 quasar pairs defined by transverse distance, redshift, and luminosity \label{tab:ranges}}
\begin{tabular}{ c | c | c }
    Parameter & Bin & Range \\
    \hline
    \multirow{3}{*}{\shortstack{{\bf Transverse distance} \\ [2ex] $r_{\perp}\ [{\rm Mpc}/h]$}}
      & Low    & $0.00$--$0.67$ \\
      & Mid    & $0.67$--$1.33$ \\
      & High   & $1.33$--$2.00$ \\
    \hline
    \multirow{3}{*}{\shortstack{{\bf Redshift} \\ [2ex] $z_{\rm f/g}$}} 
      & Low    & $2.09$--$2.22$ \\
      & Mid    & $2.22$--$2.40$ \\
      & High   & $2.40$--$3.63$ \\
    \hline
    \multirow{3}{*}{\shortstack{{\bf Luminosity} \\ [2ex] 
    $\log_{10}\left(\frac{\nu L_\nu|_{1500\text{\AA}}}{[\ergs]}\right)$
    }} 
      & Low    & $43.44$--$45.06$ \\
      & Mid    & $44.63$--$45.44$ \\
      & High   & $45.02$--$46.81$ \\
\end{tabular}
\end{table}
}

\subsection{\lya transmission} \label{sec:Lya-transmission}

Following previous studies \citep[][]{2008MNRAS.391.1457K,2019ApJ...884..151J}, we focus on the (normalized) transmitted flux, defined as the ratio of the observed flux to the unabsorbed continuum flux, $f_{q}/C_{q}$, to investigate variations in \lya absorption within the quasar proximity region. Since our continuum fitting procedure yields the mean expected flux, $\overline{F}C_{q}$, for each quasar spectrum, we need to estimate the mean transmission, $\overline{F}$, in order to compute the transmitted flux. 

\citet{2024ApJ...976..143T} developed a Convolutional Neural Network that predicts the unabsorbed quasar continuum based on the red side of the \lya emission line, referred to as the Lyman-$\alpha$ Continuum Analysis Network (LyCAN). They applied LyCAN to a high-S/N subset of the DESI Y1 quasars ($2.1<z<4.2$) to predict their unabsorbed continua and then measured the effective \lya optical depth, which they modeled as
\begin{eqnarray} 
    \tau_{\rm eff}(z) = \tau_{0}(1+z)^{\gamma}, 
\end{eqnarray}
with best-fit parameters $\tau_{0}=(2.46\pm0.14)\times10^{-3}$ and $\gamma = 3.62\pm0.04$. In the following, we use these best-fit values to compute the mean transmission as $\overline{F}(z)=\exp[-\tau_{\rm eff}(z)]$.

Our proximity measurements are evaluated relative to a redshift- and S/N-matched control sample constructed from the same \lya forest catalog (Section~\ref{sec:control_sample}). This design mitigates continuum systematics shared between the proximity and control samples; in particular, smooth multiplicative continuum errors that vary slowly with wavelength should similarly impact the proximity and control samples. The continuum fitting is also performed on the scale of entire \lya forest skewers (typically several hundred $\hinvmpc$), which is much larger than the proximity region adopted here (e.g., Figure~\ref{fig:qso_pair_spectrum}). We therefore do not expect these smooth continuum systematics to generate localized features confined to the proximity region that could mimic the differential trends among our subsamples.

Solid lines in the right panels of Figure~\ref{fig:sub_all} represent the stacked \lya transmission, $f_{q}/C_{q}$, for background quasars in each subsample bin as a function of the line-of-sight separation from the foreground quasars, with a bin width of $3\hinvmpc$. We also show the stacked \lya transmission for the corresponding control samples as dashed lines for reference. Each color corresponds to the one shown in the left panel. The error bars indicate the standard error of the mean in each pixel bin. We note that the errors in the control samples are smaller than those in the background quasars by approximately a factor of $1/\sqrt{10}$, which is expected since each quasar pair is assigned 10 control sample quasar pairs. We emphasize that even after dividing the sample into subsamples, the error bars remain small enough to allow for meaningful comparisons between different subsample bins.

\noindent \textbf{Transverse-distance subsample:} Let us first examine the results for the transverse-distance subsample in panel (a2) of Figure~\ref{fig:sub_all}. We can clearly see that \lya absorption in the background quasar spectra is enhanced around the foreground quasars, with a stronger signature for sightlines closer to the foreground quasars. The difference between $r_{\perp}=0.00{\rm -}0.67\hinvmpc$ and $r_{\perp}=0.67{\rm -}1.33\hinvmpc$ is more marked than the difference between $r_{\perp}=0.67{\rm -}1.33\hinvmpc$ and $r_{\perp}=1.33{\rm -}2.00\hinvmpc$. On the other hand, the \lya transmissions for the control samples in the three subsample bins are consistent with each other, as expected. We note that the differences in error sizes reflect variations in the number of quasar pairs in each subsample bin. The excess absorption in proximate quasar pairs suggests that the clustering of \ion{H}{1} in the dense environment around foreground quasars has a greater impact on \lya absorption than their ionizing radiation. We also find a small dip at $r_{\parallel} \simeq -20\hinvmpc$, particularly in the innermost bin (black solid line), which aligns with the expected position of SiIII absorption at the rest-frame wavelength of $1206.5\text{\AA}$ (vertical pink dotted line).

\noindent \textbf{Redshift subsample:} Panel (b2) of Figure~\ref{fig:sub_all} illustrates the redshift dependence of \lya transmission in the quasar proximity region. We do not find a clear difference between the foreground quasar redshift bins, apart from variations in the mean transmission, $\overline{F}(z)$. The transmission of the control samples exhibits a slight gradient from left to right (i.e., from the near side to the far side), likely reflecting the same redshift evolution of $\overline{F}(z)$ mentioned above. This trend is not unique to the redshift subsamples but is also observed in the other two subsample categories.

\noindent \textbf{Luminosity subsample:} In panel (c2) of Figure~\ref{fig:sub_all}, we present the results for the luminosity subsamples. Interestingly, we do not find any dependence of the \lya transmission on the foreground quasar luminosity, even though the luminosity in the bins differs by nearly an order of magnitude. We will further investigate this feature in the following sections. The only exception is that the position of the dip in the transmission at $r_{\parallel} \simeq 0$ for the highest luminosity bin is slightly shifted with respect to the other two subsamples toward the far side (i.e., by approximately one bin width, $3\hinvmpc$), compared to those for the other two bins. In the context of the \lya forest–quasar cross-correlation, this offset is commonly parameterized as a free parameter, $\Delta r_{\parallel}$, representing a small systematic bias in the quasar redshift estimates. The baryon acoustic oscillation (BAO) analysis using the same DESI Y1 \lya forest data as in this paper found a best-fit value of $\Delta r_{\parallel} = 0.077^{+0.061}_{-0.062}\hinvmpc$ from the cross-correlation alone \citep{2025JCAP...01..124A,2023AJ....166...66B,2025JCAP...01..130B}. The results here suggest that the bias could depend on luminosity.

A plausible explanation for such a luminosity-dependent shift is the well-known blueshift of the high-ionization \ion{C}{4} emission line in luminous quasars \citep[e.g.,][]{1982ApJ...263...79G,2001AJ....122..549V,2023MNRAS.523..646T}, which can lead to an underestimation of their systemic redshift and hence produce an apparent displacement of associated absorption features toward the far side. The typical magnitude of this blueshift corresponds to a few hundred up to $\sim 1000\kms$ (i.e., of order $10\hinvmpc$ at $z\sim2.5$). In practice, however, {\tt Redrock}, the DESI redshift pipeline, determines redshifts by fitting the full quasar spectrum with multiple emission features, so the effect is not a direct one-to-one shift from \ion{C}{4} alone, but rather contributes to a residual systematic uncertainty in the estimated quasar redshifts. The observed offset of $\sim 3\hinvmpc$ is well within the expected scale of such blueshift-induced biases, suggesting that the contribution from \ion{C}{4} systematics provides a reasonable explanation. The transmission spectra shown in the right panels also appear to show an asymmetric profile, such that the transmission is higher on the far side ($r_{\parallel} > 0$) than the near side of the foreground quasar. We return to this point in Section~\ref{sec:quasar_emission}.

\section{Interpretation} \label{sec:interpretation}

Following the results on the \lya transmission presented in the previous section, we further investigate the properties of \ion{H}{1} absorption in the quasar proximity region, with a particular focus on the impact of quasar radiation.

\subsection{Optical depth and Quasar radiation}

As discussed in Section~\ref{sec:Lya-transmission}, our measurements of \lya transmission suggest that the effect of \ion{H}{1} gas clustering in the dense environments around quasars dominates over the extra ionization caused by their radiation. We now proceed to investigate these two contributions more quantitatively. To connect the overdensity and ionization state of \ion{H}{1} gas clouds to our measurements, we focus on the optical depth,
\begin{eqnarray}
    \tau = -\ln(f_{q}/C_{q}),
\end{eqnarray}
which is proportional to the neutral hydrogen number density. We adopt the notation $\tau_{\rm prx}$ for the optical depth measured from the background quasars, which probe the proximity regions of their corresponding foreground quasars, and $\tau_{\rm IGM}$ for the optical depth measured from the control samples that represent the average IGM. Following previous studies \citep[e.g.,][]{2005MNRAS.361.1015R,2008ApJ...673...39F}, we introduce a simple relation between $\tau_{\rm prx}$ and $\tau_{\rm IGM}$, expressed in terms of the density contrast of \ion{H}{1} gas clouds and the ionizing photon flux from foreground quasars.

\subsubsection{A simple model for optical depth}

In the redshift range of $z = 2$–3.5 relevant to our foreground quasars, the neutral hydrogen fraction in the IGM is known to be extremely low — for instance, $x_{\rm H\,\textsc{i}} \lesssim 3 \times 10^{-5}$ at $z < 6$ \citep[][]{2016ARA&A..54..313M}. Moreover, theoretical considerations by \citet{2001ApJ...559..507S} indicate that typical \lya forest absorbers have neutral hydrogen column densities of $N_{\rm H\,\textsc{i}} \lesssim 10^{14}{\rm cm^{-2}}$ at $z < 5$, significantly below the Lyman limit of $\sim 10^{17}{\rm cm^{-2}}$, above which the gas becomes optically thick to ionizing photons. This implies that the IGM is largely transparent to ionizing radiation in most regions. In our analysis, this assumption is further justified as we exclude systems with high column densities, such as DLAs, by applying appropriate masks in the construction of the \lya forest catalog.\footnote{If DLAs were statistically more abundant in the vicinity of foreground quasars, our masking procedure could lead to an underestimation of the neutral hydrogen content. In addition, while residual contamination from undetected DLAs may remain, we do not explicitly model this effect in our analysis.} In such a highly ionized and optically thin environment, assuming photoionization equilibrium, the following relation holds:
\begin{eqnarray}
    \Gamma_{\rm tot}n_{\rm H\,\textsc{i}} 
        = \alpha_{\rm A}(T)n_{\rm e}n_{\rm H\,\textsc{ii}} \propto \alpha_{\rm A}(T)\Delta^{2},  \label{eq:ion_equil}
\end{eqnarray}
where $n_{\rm H\,\textsc{i}}$, $n_{\rm H\,\textsc{ii}}$, and $n_{\rm e}$ are the number densities of neutral hydrogen, ionized hydrogen, and electrons, respectively. The total hydrogen number density is given by $n_{\rm H} = n_{\rm H\,\textsc{i}} + n_{\rm H\,\textsc{ii}}$, and $\Delta$ denotes the gas overdensity, which is approximately equal to the hydrogen overdensity, $\rho_{\rm H}/\bar{\rho}_{\rm H}$. $\Gamma_{\rm tot}$ is the total photoionization rate, and $\alpha_{\rm A}(T)$ is the case-A recombination coefficient for optically thin gas at temperature $T$, which is well approximated by the fitting formula $\alpha_{\rm A}(T) \simeq 4.13\times10^{-13}(T/10^{4}\K)^{-0.71}\cubiccms$ \citep{Draine}. Note that we used $n_{\rm H\,\textsc{ii}} \simeq n_{\rm H}$ and $n_{\rm e} \simeq n_{\rm H} + 2n_{\rm He} \propto n_{\rm H}$, where $n_{\rm He}$ is the total helium number density, valid under the assumption of fully ionized hydrogen and helium gas. In our case, we can reasonably assume that the proportionality constant in Equation~\ref{eq:ion_equil} is independent of local physical conditions, such as gas temperature or overdensity, as long as the contribution of helium to the electron number density is spatially uniform.\footnote{While the reionization of neutral helium (\ion{He}{1}), which likely proceeded similarly to that of neutral hydrogen, had already completed by the relevant redshifts \citep[e.g.,][]{2006ARA&A..44..415F}, \ion{He}{2} reionization is thought to occur around $z \gtrsim 3$ \citep[e.g.,][]{2005SSRv..116..625C,2009ApJ...694..842M,2023RAA....23d5007Y}. Nevertheless, incomplete knowledge of the \ion{He}{2} reionization state within local IGM gas can introduce up to an $8\%$ uncertainty in the estimate of the electron number density \citep[e.g.,][]{2012ApJ...746..125H}.}

According to Equation~\ref{eq:ion_equil}, both optical depths, $\tau_{\rm IGM}$ and $\tau_{\rm prx}$, can be related to the photoionization rate and gas overdensity. In the IGM control samples, both quantities follow simple assumptions: photoionization is driven solely by the uniform UVB ($\Gamma_{\rm UVB}$), and the gas overdensity is unity on average across many sightlines, as defined by our continuum fitting procedure. Under these conditions, the optical depth is given by $\tau_{\rm IGM} \propto C_{\rm H\,\textsc{ii}} \alpha_{\rm A}(T) / \Gamma_{\rm UVB}$, where $C_{\rm H\,\textsc{ii}}\equiv\langle n^{2}_{\rm H\,\textsc{ii}} \rangle/\langle n_{\rm H\,\textsc{ii}} \rangle^{2}$ is the clumping factor that accounts for small-scale inhomogeneities in the ionized hydrogen density. In contrast, in quasar proximity regions, where the total photoionization rate includes a local contribution from foreground quasars ($\Gamma_{q}$), the \ion{H}{1} optical depth is
$\tau_{\rm prx} \propto C_{\rm H\,\textsc{ii}}\alpha_{\rm A}(T)\Delta^{2} / (\Gamma_{\rm UVB} + \Gamma_{q})$. In this case, $\Delta$ remains significantly different from unity even after stacking quasar pairs, reflecting the typically overdense environments surrounding quasars. Note that although we assume the clumping factor to be the same in the IGM and in quasar proximity regions, it has been shown that local clumping factors can vary with the local gas overdensity \citep[e.g.,][]{2015ApJ...810..154K}. In this case, we would need to consider the variation of the clumping factor in proximity regions separately from the gas overdensity. However, for simplicity, we neglect such variation in this study.

Assuming a simple temperature–density relation, $T \propto \Delta^{\gamma - 1}$, and using the above fitting formula for the recombination coefficient, the ratio of the two optical depths can be written as
\begin{eqnarray}
    \frac{\tau_{\rm prx}}{\tau_{\rm IGM}} = \frac{\Delta^{2-0.71(\gamma-1)}}{(\Gamma_{\rm UVB}+\Gamma_{q})/\Gamma_{\rm UVB}}. \label{eq:tau_ratio}
\end{eqnarray}
\citet{2012ApJ...757L..30R} analyzed a sample of more than 5000 \ion{H}{1} absorbers in the IGM over the redshift range $2.0 < z < 2.8$ and inferred a power-law index of $\gamma - 1 = 0.46 \pm 0.05$ for the temperature–density relation at a mean redshift of $\bar{z} = 2.4$. In the following analysis, we adopt this value for $\gamma$. The most important source of inaccuracy in Equation~\ref{eq:tau_ratio} is that $\tau_{\rm prx}$ is measured from the flux averaged over a DESI resolution element, and averaging flux is not the same as averaging optical depth because of the nonlinear relation between them. Similarly, $\tau_{\rm IGM}$ is defined from the mean transmitted flux of the IGM rather than an average of optical depth. Thus, Equation~\ref{eq:tau_ratio} only applies strictly in the limit that all optical depths are $\tau \ll 1$. Nevertheless, it provides a useful approximate tool for interpreting our measurements in terms of density profiles around quasars, as a prelude to full forward modeling from IGM simulations.

\subsubsection{Quasar ionization} \label{sec:qso_ionization}

\begin{figure*}[ht!]
    \centering
    \includegraphics[width=\linewidth]{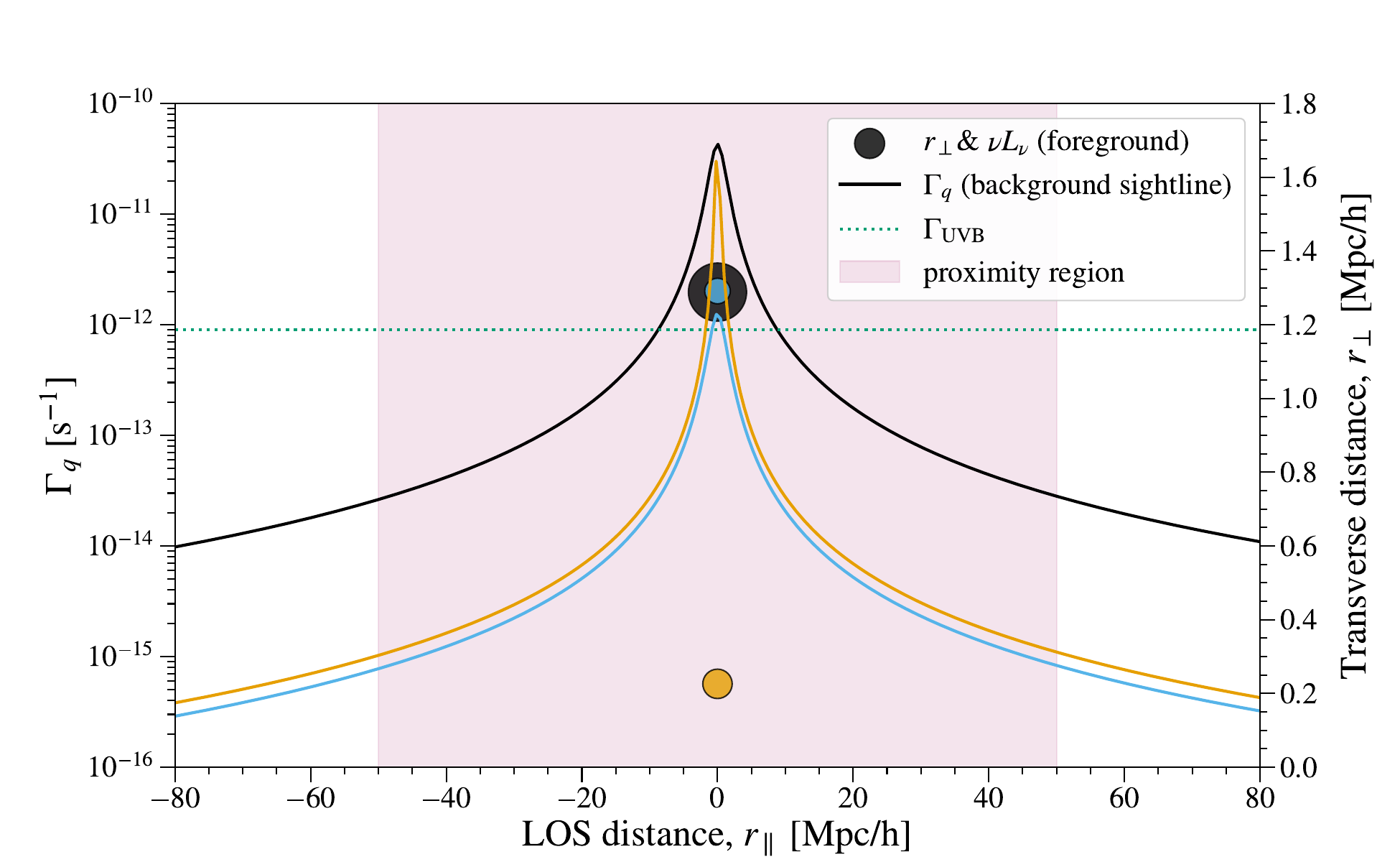}
\caption{Ionization rates due to three foreground quasars at $z\simeq2.2$ (circle) on their background quasar sightlines (solid line), compared with the UVB ionization rate (green dotted line). Similar to Figure~\ref{fig:qso_pair_spectrum}, but with circle size reflecting the luminosity of the foreground quasar: $\log(\nu L_\nu|_{1500\text{\AA}}/[\ergs]) =$ 46.39 (black), 44.98 (orange), and 44.86 (sky blue).
\label{fig:qso_ionization_rate}}
\end{figure*}

Throughout this work, we adopt a UVB ionization rate of $\Gamma_{{\rm UVB}} = 9 \times 10^{-13}\ \rm{s}^{-1}$ over the redshift range $2\lesssim z\lesssim 3$ \citep[][]{2012ApJ...746..125H}. On the other hand, since the luminosities of the foreground quasars are known, we can estimate the photoionization rate at each pixel along the background quasar spectra, $\Gamma_{q}$. To do this, we extrapolate the quasar ionizing photon flux from the luminosity measured at rest-frame 1500\AA, $L_\nu|{1500\text{\AA}}$. Assuming typical values for the quasar continuum spectral slope \citep{2015MNRAS.449.4204L},
\begin{eqnarray}
    \alpha = 
    \begin{cases}
        -0.61 \quad (\nu < \nu_{\rm LL}) \\
        -1.70 \quad (\nu > \nu_{\rm LL}), 
    \end{cases}
\end{eqnarray}
the quasar luminosity density, $L_{q}(\nu_{q})$ (in $\rm{erg}\ \rm{s}^{-1}\ \rm{Hz}^{-1}$), can be expressed as
\begin{eqnarray}
    L_{q}(\nu_{q}) = L_{q,\nu} |_{1500\text{\AA}} \left(\frac{\nu_{\rm LL}}{\nu |_{1500\text{\AA}}} \right)^{-0.61} \left(\frac{\nu_{q}}{\nu_{\rm LL}}\right)^{\alpha}, 
\end{eqnarray}
where $\nu_q$ is the rest-frame frequency at the quasar's position, and $\nu_{\rm LL}$ is the Lyman limit frequency. Assuming isotropic emission, the quasar flux per unit frequency at a pixel in the background quasar spectrum, denoted as $F_q(\nu)$ (in $\rm{erg}\ \rm{cm}^{-2}\ \rm{s}^{-1}\ \rm{Hz}^{-1}$), is given by 
\begin{eqnarray} 
    F_q(\nu) = (1+z_{\rm rel}) \frac{L_q\left( (1+z_{\rm rel})\nu \right)}{4\pi d_{L, \rm rel}^2}. 
\end{eqnarray} 
Here, $z_{\rm rel}$ is the relative redshift between the foreground quasar and the pixel, defined by $1+z_{\rm rel} = a(\eta_{\rm pix})/a(\eta_{\rm emi})$, where $\eta_{\rm emi}$ denotes the conformal time when the ionizing photons were emitted from the foreground quasar. The conformal time $\eta_{\rm pix}$ corresponds to the comoving distance $\chi_{\rm pix}$ to the pixel and is given by $\eta_{\rm pix} = \eta_0 - \chi_{\rm pix}$, with $\eta_0$ being the conformal time at the present epoch. The relative luminosity distance is denoted by $d_{L, \rm rel}$ and calculated accordingly \citep[see][for more details]{2024MNRAS.531.2912H}.

We now briefly discuss the comoving separation between the foreground quasar and the pixel, $\chi_{q\text{-}{\rm pix}} = \eta_{\rm pix} - \eta_{\rm emi}$. In practice, quasars have finite lifetimes, meaning there is a maximum distance up to which their radiation can reach. This sets a lower bound on $\eta_{\rm emi}$ and consequently an upper bound on $\chi_{q\text{-}{\rm pix}}$. Ideally, we should assess whether each pixel lies within the illuminated region, and set the quasar flux to zero for those beyond it. However, for simplicity, we assume that all foreground quasars have infinite lifetimes and that all pixels of background quasars are exposed to the quasar radiation. We note that this assumption is intentionally simplified, as quasars are expected to have finite lifetimes and anisotropic emission. 
These effects are discussed in Section~\ref{sec:quasar_emission}.

Finally, the ionization rate contributed by the foreground quasar can be written as: 
\begin{eqnarray}
    \Gamma_{q} &=& \int_{\nu_{\rm LL}}^{\infty} \mathrm{d}\nu \frac{F_{q}(\nu)}{h\nu} \sigma_{\rm ion}(\nu) \label{eq:gamma_q_i} 
\end{eqnarray} 
where $\sigma_{\rm ion}(\nu)$ represents the photoionization cross section for \ion{H}{1} atoms, given by 
\begin{eqnarray} 
    \sigma_{\rm ion} = \sigma_{\nu_{\rm LL}} \left(\frac{\nu}{\nu_{\rm LL}}\right)^{-3}, \label{eq:sigma_pi} 
\end{eqnarray} 
with $\sigma_{\nu_{\rm LL}}=6.3 \times 10^{-18} \ \rm{cm}^2$ being the cross section at the Lyman limit \citep{Osterbrock, Draine}.

\begin{figure*}[ht!]
\centering
    \includegraphics[width=\textwidth]{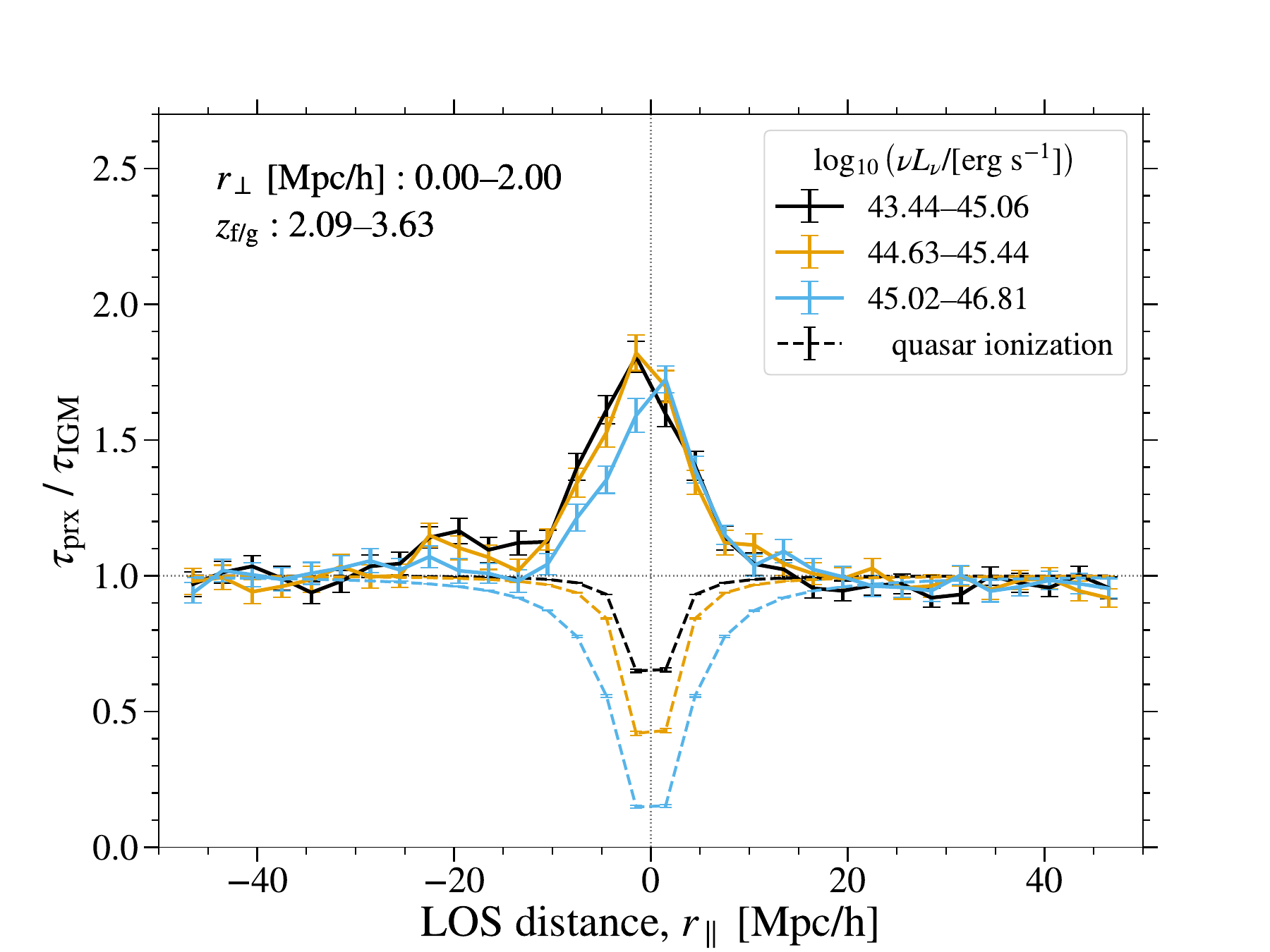}
\caption{Luminosity dependence of \lya optical depth variations in quasar proximity regions (solid lines). For comparison, we also show the expected contribution from quasar ionizing radiation alone (dashed lines), corresponding to the case where the overdensity is fixed at $\Delta=1$ in Equation~\ref{eq:tau_ratio}. Line colors and vertical lines are the same as in Figure~\ref{fig:sub_all}. \label{fig:tau_ratio_luminosity}}
\end{figure*}

As expected from its definition, the ionization rate $\Gamma_{q}$ scales with the quasar luminosity and inversely with the square of the distance to the pixels. To illustrate this behavior, in Figure~\ref{fig:qso_ionization_rate}, we present the ionization rates along the background quasar sightlines for three quasar pairs, each representing different combinations of low- and high-luminosity foreground quasars, as well as small and large separations. The black curve shows a pair with a luminous foreground quasar and a large separation, illustrating that the ionization rate from the foreground quasar exceeds that of the UVB around the proximity region. The orange curve presents the opposite case, featuring a faint foreground quasar with a small separation, where the peak ionization rate is comparable but the region of excess $\Gamma_{q}$ is much narrower. In sky blue, we show the remaining case of a faint foreground quasar with a large separation, where $\Gamma_{q} \simeq \Gamma_{\rm UVB}$ even at the peak. This comparison highlights that, for typical luminosities in our sample ($\log(\nu L_\nu|_{1500\text{\AA}}/[\ergs]) \sim 45$), the quasar photoionization rate falls below the UVB level beyond $\sim 2\hinvmpc$, consistent with our choice to limit the analysis to these separations (see Section~\ref{sec:qso_pairs}).

\begin{figure*}[ht!]
    \begin{minipage}[t]{0.54\linewidth}
        \centering
        \includegraphics[width=\linewidth]{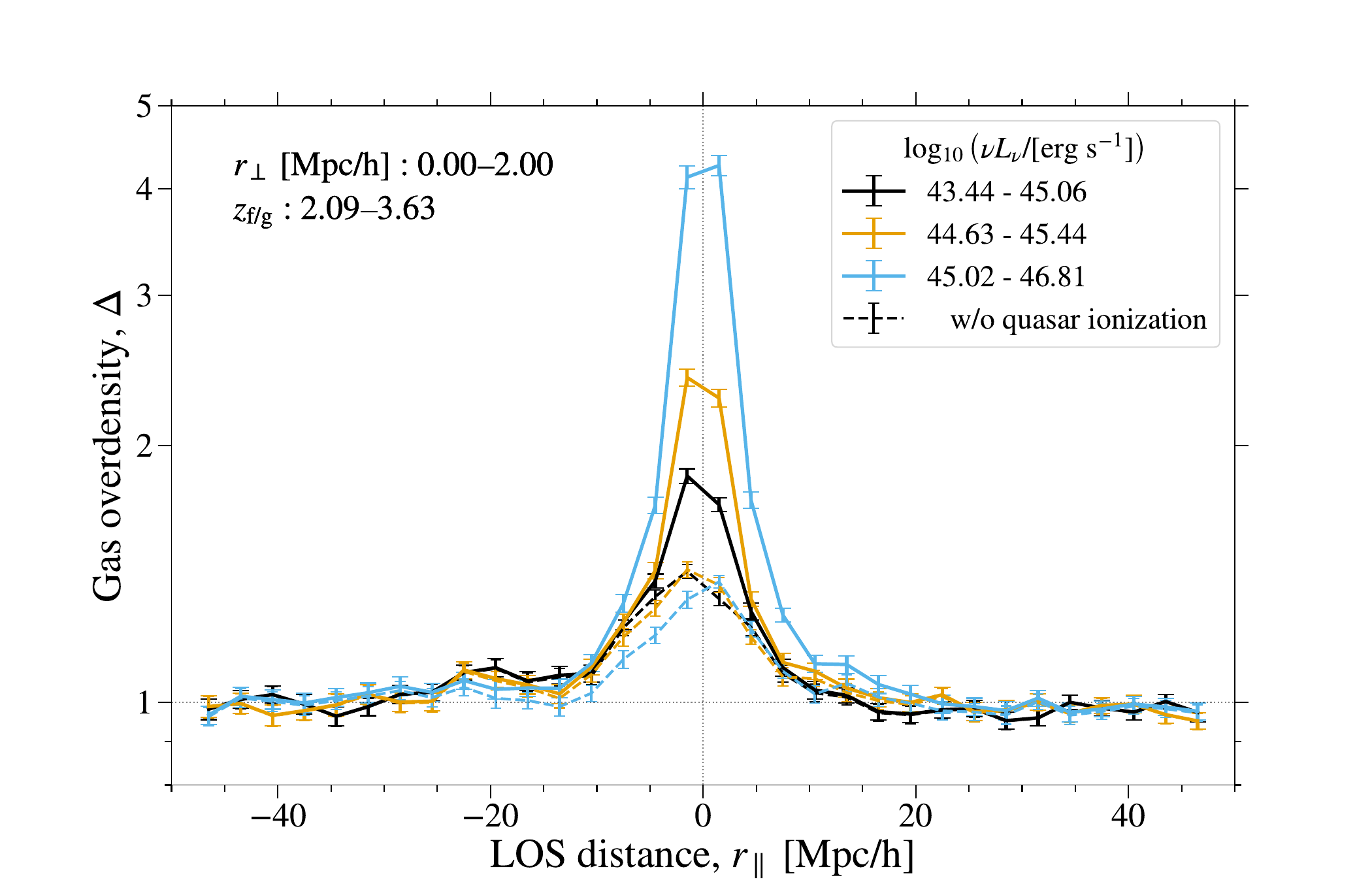} 
    \end{minipage}
    \hspace{-0.05\linewidth}
    \begin{minipage}[t]{0.46\linewidth}
        \centering
        \includegraphics[width=\linewidth]{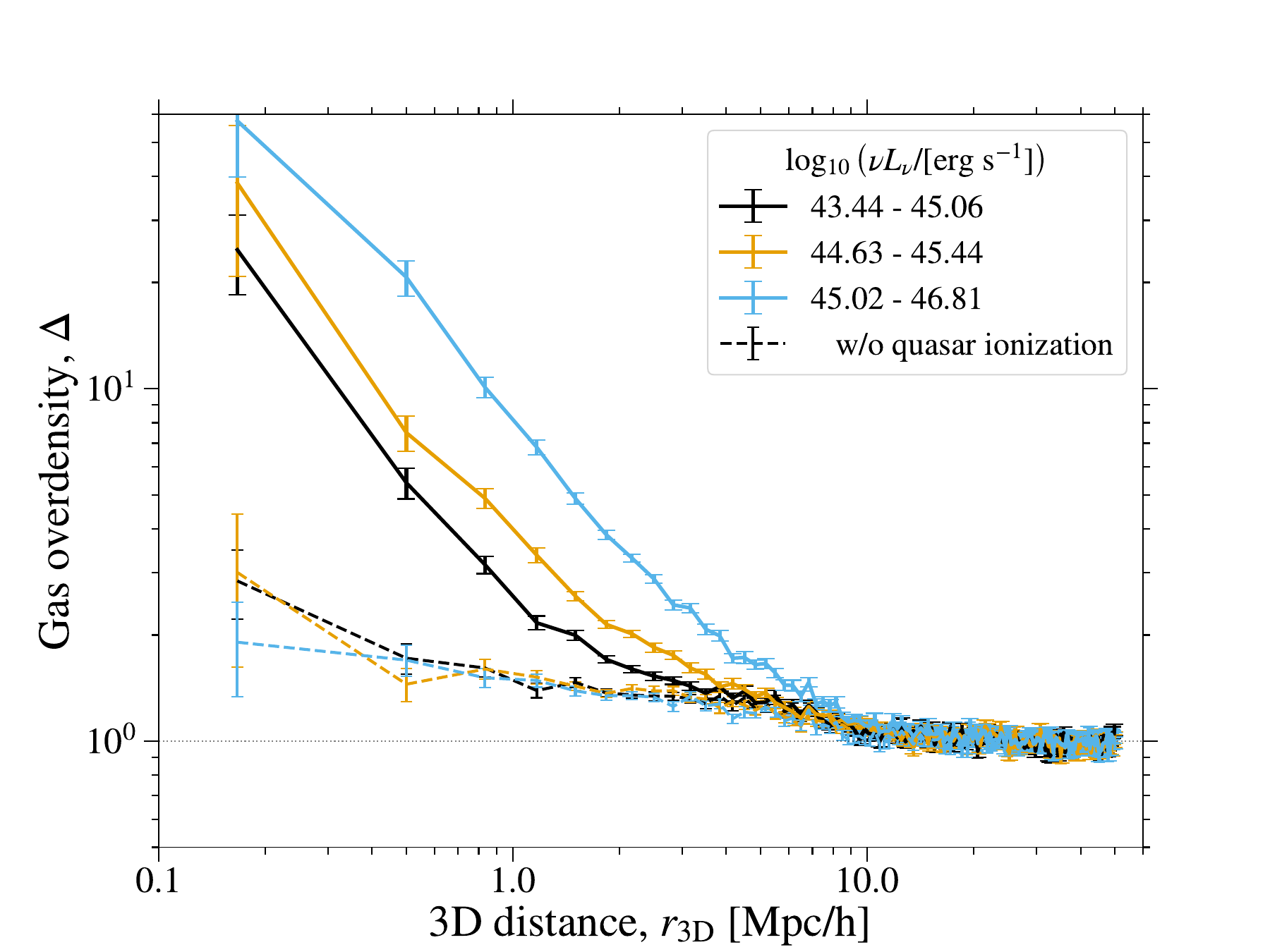}
    \end{minipage}
\caption{Inferred gas overdensity profiles around quasars for the luminosity subsamples, shown as a function of line-of-sight separation $r_{\parallel}$ (left panel) and 3D distance $r_{\rm 3D}$ (right panel). The overdensities are derived from the $\tau_{\rm prx}/\tau_{\rm IGM}$ profiles in Figure~\ref{fig:tau_ratio_luminosity} using Equation~\ref{eq:tau_ratio}, with (solid lines) and without (dashed lines) including the contribution from quasar ionizing radiation. Line colors and vertical lines are the same as in Figure~\ref{fig:sub_all}.
\label{fig:sub_luminosity_r3d}}
\end{figure*}

\subsubsection{Observed optical depth}

Based on the discussion above, we now compare our measurements with the model prediction for the \lya optical depth. In Figure~\ref{fig:tau_ratio_luminosity}, the solid lines show the ratio of the observed optical depths, $\tau_{\rm prx}/\tau_{\rm IGM}$, for the three luminosity subsamples defined in Figure~\ref{fig:sub_all}. The dashed lines in the figure indicate the expected contribution from quasar ionizing radiation alone, corresponding to the case where the overdensity is set to $\Delta = 1$ in Equation~\ref{eq:tau_ratio}: 
\begin{eqnarray} 
    \left[\frac{\tau_{\rm prx}}{\tau_{\rm IGM}}\right]_{\rm rad} \equiv \left. \frac{\tau_{\rm prx}}{\tau_{\rm IGM}}\right|_{\Delta=1} =\frac{\Gamma_{\rm UVB}}{\Gamma_{\rm UVB}+\Gamma_{q}}, 
\end{eqnarray}
where, for each quasar pair, $\Gamma_{q}$ is calculated based on the luminosity of the foreground quasar and the positions of pixels along the background quasar spectrum, following the equations presented in Section~\ref{sec:qso_ionization}. This prediction implicitly assumes 
isotropic and steady emission of the foreground quasar.

We estimated the mean and its uncertainty in each line-of-sight separation bin (with a width of $3\hinvmpc$) using bootstrapping, for both the observed optical depth ratio and its expected contribution from quasar radiation. Specifically, we generated 100 bootstrap samples by randomly resampling quasar pairs with replacement, and computed the mean and standard error of the optical depth ratio from these resampled datasets. We compare the error estimation methods for the transmission of the \lya forest - standard deviation versus bootstrapping - in Appendix~\ref{error_comparison}, and find that our bootstrapping approach provides sufficient precision for comparing different subsamples and for comparing the data with model predictions, given that high-precision error estimation is not required. We estimate the mean and its uncertainty for all subsequent measurements using the bootstrap method described above.

As expected from the results on \lya transmission, we find no significant differences in the observed optical depths among the luminosity subsamples, except for a shift in the peak position. The enhanced optical depth near foreground quasars suggests that the clustering of \ion{H}{1} absorbers in dense environments has a greater impact than the quasar radiation. Consistent with this picture, previous TPE studies also reported excess \ion{H}{1} absorption around quasars, pointing to overdense environments \citep[e.g.,][]{2013ApJ...776..136P,2019ApJ...884..151J}. On the other hand, the expected contribution from quasar radiation is substantial enough to alter the optical depth ratio and shows a clear dependence on quasar luminosity. Similar effects were anticipated in simulations by \citet{2004ApJ...610..642C}, who showed that finite lifetimes and light-cone effects can shape quasar light echoes and modify the \lya forest absorption, highlighting the sensitivity of the proximity effect to quasar radiation models. The shift in the peak position corresponds directly to the displacement of the dip in the \lya transmission. While a luminosity-dependent \ion{C}{4} blueshift provides a plausible explanation, we will also discuss in Section~\ref{sec:quasar_emission} alternative mechanisms that could produce such an offset.

The contribution from quasar radiation should be regarded as an upper limit, due to several simplifying assumptions we have made. First, as noted below Equation~\ref{eq:tau_ratio}, the optical depth of high-column-density absorbers may be underestimated due to the finite pixel size, which can obscure the full extent of the quasar ionization effect. \citet{2019ApJ...884..151J} used synthetic spectra generated from hydrodynamical simulations of the IGM to demonstrate that the observationally inferred $\Gamma_q$ tends to underestimate the true value. In addition, Lyman limit systems (LLSs), along with \ion{H}{1} gas clouds, are known to cluster in the dense environments around quasars \citep[e.g.,][]{2013ApJ...776..136P}, which can further attenuate the quasar ionizing flux reaching the pixels.
Nevertheless, it remains important to consider how to reconcile the lack of luminosity dependence in the observed optical depth with the clear luminosity dependence predicted for the contribution from quasar ionization.

\begin{figure*}[ht!]
    \begin{minipage}[t]{0.5\linewidth}
        \centering
        \includegraphics[width=\linewidth]{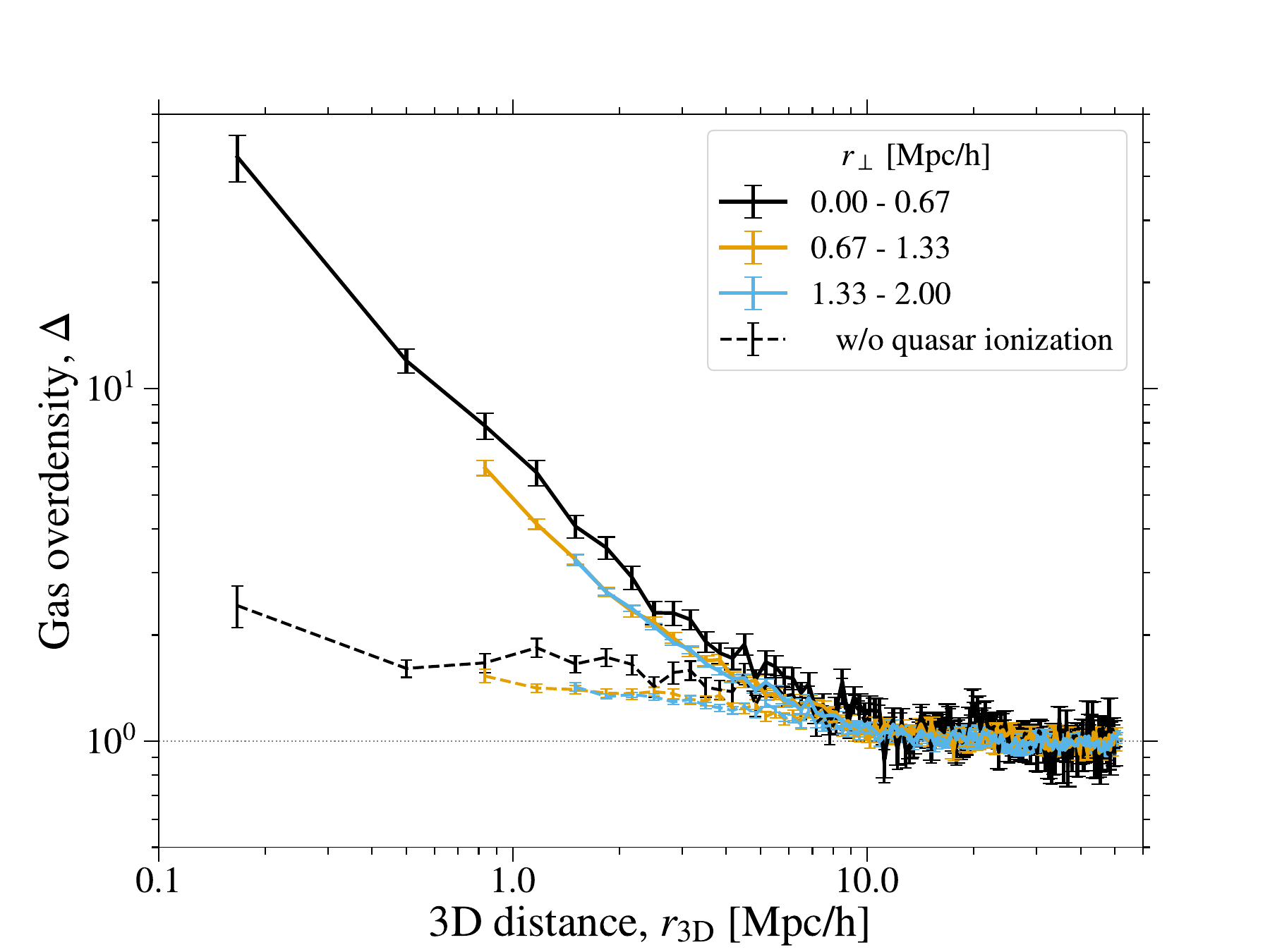} 
    \end{minipage}
    \hspace{-0.05\linewidth}
    \begin{minipage}[t]{0.5\linewidth}
        \centering
        \includegraphics[width=\linewidth]
        {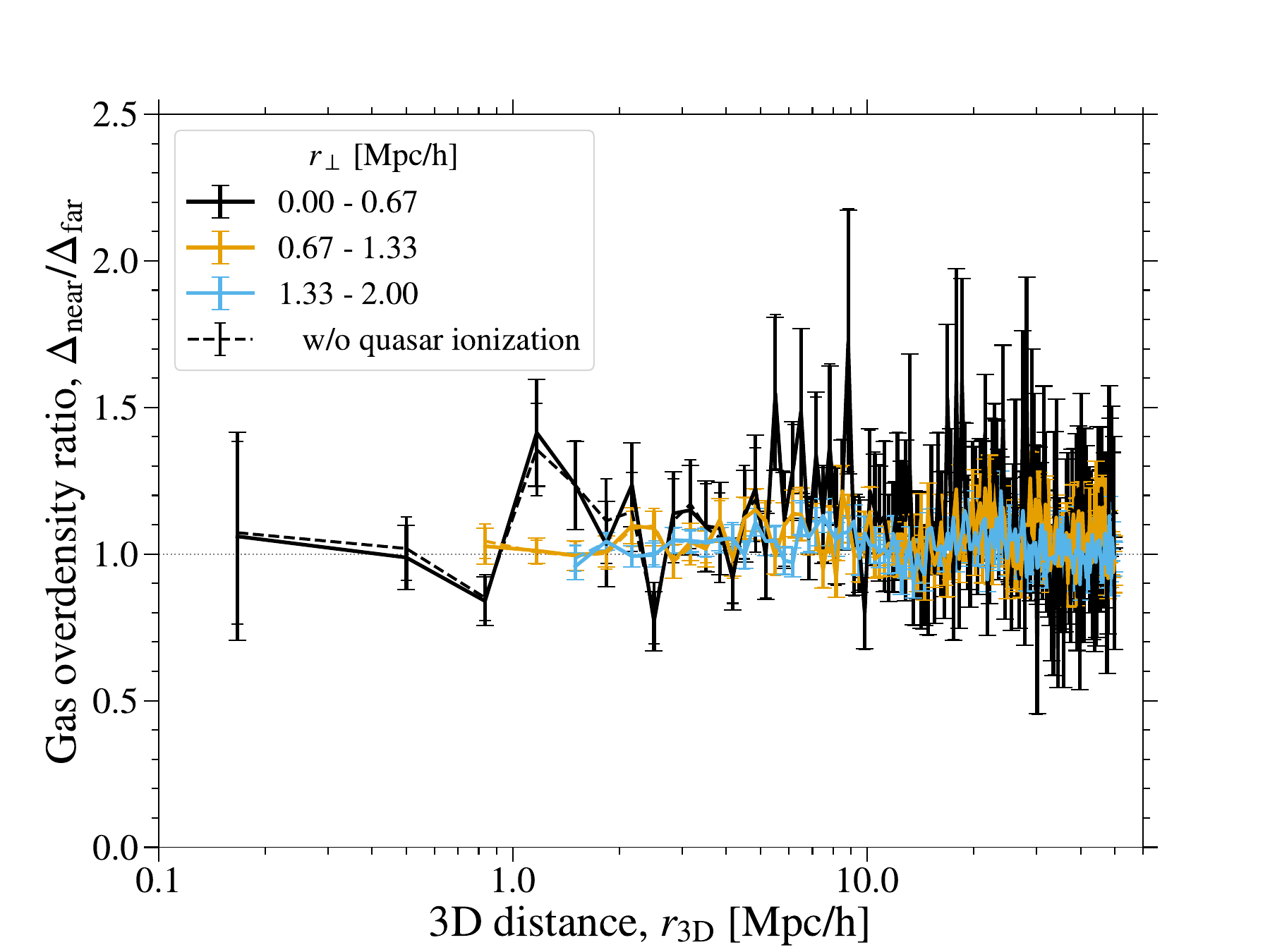}
    \end{minipage}
\caption{Inferred gas overdensity profiles around quasars for the transverse-distance subsamples (left panel), and the ratio of overdensities between the near side ($r_{\parallel} < 0$) and the far side ($r_{\parallel} > 0$) based on those profiles (right panel). The overdensities are derived from the $\tau_{\rm prx}/\tau_{\rm IGM}$ profiles for the transverse-distance subsamples using Equation~\ref{eq:tau_ratio}, with (solid lines) and without (dashed lines) including the contribution from quasar ionizing radiation. Line colors are the same as in Figure~\ref{fig:sub_all}.
\label{fig:sub_xvers_r3d}}
\end{figure*}

\subsection{Gas overdensity profile}

The contribution predicted from quasar ionization alone is given by Equation~\ref{eq:tau_ratio} with $\Delta=1$ and we interpret the difference between the predicted and measured values (dashed and solid lines in Figure~\ref{fig:tau_ratio_luminosity}) as arising from the gas overdensity. In other words, by substituting the measured optical depth ratio into the left-hand side of Equation~\ref{eq:tau_ratio}, we can estimate the average value of $\Delta$ as a function of distance.

Figure~\ref{fig:sub_luminosity_r3d} shows the gas overdensity profiles for the luminosity subsamples, plotted as a function of line-of-sight separation $r_{\parallel}$ (left; bin width $3\hinvmpc$) and three-dimensional distance $r_{\rm 3D} = (r_{\parallel}^2 + r_{\perp}^2)^{1/2}$ (right; bin width $0.33\hinvmpc$).\footnote{A finer bin width of $0.33\hinvmpc$ was adopted for $r_{\rm 3D}$, since the original $3\hinvmpc$ bin (used for $r_{\parallel}$) would be too coarse relative to the $0.67\hinvmpc$ bin size used in defining the transverse-distance subsamples, making it difficult to compare results across different transverse-distance bins.} Solid and dashed lines indicate the overdensity profiles with and without the contribution from quasar radiation, respectively. In the absence of quasar ionization ($\Gamma_{q}=0$), the three dashed profiles overlap, consistent with the expectation that the overdensity is described by a simple power-law function of the optical depth ratio, independent of luminosity. By contrast, the separation of the solid lines demonstrates the luminosity dependence of the quasar ionization rate.

We find that the gas overdensity is systematically higher around more luminous foreground quasars, and the dense environments are particularly prominent at scales $r_{\rm 3D} \lesssim 10\hinvmpc$. Similar conclusions have been drawn from LPE studies at redshifts $z=2{\rm -}4$, which analyzed the optical depth distribution along quasar sightlines. For example, \citet{2005MNRAS.361.1015R} inferred overdensities of order a few within $\sim10\hinvmpc$ to explain the observed LPE, while \citet{2007MNRAS.377..657G} reported overdensities up to $\sim5$ (or $\sim2$) within $\sim3$ (or $10$) $\hinvmpc$, and further noted a positive correlation between quasar luminosity and overdensity. These findings are qualitatively consistent with our TPE analysis, although they are based on different sightline geometries. 

Based on this result, one possible interpretation for the lack of luminosity dependence in the optical depth is that more luminous quasars reside in denser environments, and that the environmental dependence cancels out the effect of increased quasar ionizing radiation. However, such an almost exact cancellation appears coincidental. Moreover, previous measurements of quasar clustering in redshift surveys suggest that quasar clustering does not depend on luminosity \citep{2005MNRAS.356..415C,2015MNRAS.453.2779E,2017JCAP...07..017L}, or at most shows only a weak luminosity dependence \citep{2013ApJ...778...98S}, which in turn implies that quasars of different luminosity reside in halos of similar mass and environment.
These tests of luminosity dependence are anchored by relatively luminous quasars, e.g., the BOSS luminosity-binned analysis \citep{2015MNRAS.453.2779E}, whose faint-end limit ($L_{\rm bol}\ge 3.2\times10^{45}\,{\rm erg\,s^{-1}}$) roughly corresponds to $\log_{10}(\nu L_\nu|_{1500\text{\AA}}/{\rm erg\,s^{-1}})\gtrsim 44.9$, and this luminosity regime still overlaps our middle and high luminosity bins (Figure~\ref{fig:sub_all}; Table~\ref{tab:ranges}).
This apparent inconsistency motivates us to consider alternative explanations for the lack of luminosity dependence, focusing on aspects related to quasar emission itself. We explore these possibilities in the following subsection.

\subsection{Quasar emission: \\ variation, lifetime, and geometry \label{sec:quasar_emission}}

So far, we have assumed quasars to remain constantly luminous, shine indefinitely, and emit radiation isotropically, for simplicity. In reality, however, their luminosity can evolve over time, their lifetimes are finite, and their emission may be anisotropic. In this section, we consider how these factors affect our interpretation.

First, one plausible explanation for the lack of luminosity dependence in the \lya optical depth involves the time evolution of quasar luminosity. The absence of strong luminosity-dependent clustering implies that quasars of different luminosities reside in halos of similar mass, which suggests that quasar luminosity may not be tightly coupled to halo mass. This scenario can be understood if individual quasars undergo substantial luminosity variation, such that the most luminous quasars represent short-lived, high-luminosity phases within longer episodic lifetimes. If the duration of these luminous episodes is shorter than the light-crossing time across the transverse separation, then the observed luminosity along the line of sight may not reflect the ionizing flux seen by the transverse sightlines. In this case, the effective ionizing radiation in the transverse direction could be decoupled from the observed luminosity, naturally accounting for the lack of luminosity dependence. This interpretation does not conflict with the luminosity dependence often discussed for the LPE; in particular, that luminosity scaling has been explicitly used in UVB measurements \citep[e.g.,][]{2008A&A...491..465D,2009RvMP...81.1405M}. A similar idea is proposed by \citet{2006ApJ...641...41L}, who suggest that quasars spend significantly more time in low-luminosity states than in high-luminosity ones, thereby biasing instantaneous luminosity measurements.

Furthermore, the quasar proximity zone has a characteristic apparent boundary that is asymmetric with respect to the quasar, resulting from the effect of light-travel time. It extends farther on the near side than on the far side, and the overall size grows as the quasar ages \citep[see, for example, the figures in][for visualization]{2005ApJ...623..683Y,2019ApJ...882..165S, 2024MNRAS.531.2912H}. Based on this fact, we can propose an alternative explanation for the shift in the dip position of the transmission (or the peak position of the optical depth). Specifically, the contribution of quasar ionization is more significant in the near side of luminous quasars, leading to the suppression of the optical depth for the most luminous quasar subsample (sky blue solid line in Figure~\ref{fig:tau_ratio_luminosity}) in the near side ($r_{\parallel} < 0$).

Finally, the unified model of active galactic nuclei (AGNs) \citep[e.g.,][]{1993ARA&A..31..473A,2015ARA&A..53..365N} suggests that quasar radiation is emitted through bi-polar cones, a scenario supported by several observational studies \citep[e.g.,][]{2004Natur.430..999W,2007ApJ...655..735H,2013ApJ...776..136P}. In particular, \citet{2019ApJ...884..151J} compared the LPE and TPE, and showed that the transverse illumination is less than $\sim30\%$ of that along the line of sight, reinforcing the case for anisotropic obscuration. The \lya optical depth profiles in Figure~\ref{fig:tau_ratio_luminosity} assume isotropic emission. If, instead, quasars emit ionizing radiation anisotropically through bi-polar cones, this assumption would lead to an overestimate of the ionizing flux reaching the transverse sightlines. In such a scenario, the lack of luminosity dependence can be naturally explained by the reduced or absent ionizing flux in directions perpendicular to the radiation cones. Consequently, the quasar ionization contribution shown in the same figure (dashed lines) likely exceeds the actual effect, potentially exaggerating the apparent difference between luminosity subsamples. In fact, if the transverse sightlines lie outside the radiation cones, the inferred gas overdensity profiles — after removing the overestimated radiation contribution (e.g., setting $\Gamma_{q}=0$) — become consistent across luminosity bins (dashed lines in the right panel of Figure~\ref{fig:sub_luminosity_r3d}), as the overdensity profile then depends solely on the optical depth. This scenario also helps resolve the apparent inconsistency with the lack of luminosity dependence in quasar clustering.
 
While the discussion so far has focused on how quasar emission properties may explain the lack of luminosity dependence in the \lya optical depth, it is also instructive to examine the dependence on transverse distance. Both finite quasar lifetimes and anisotropic emission can imprint characteristic spatial signatures on the three-dimensional distribution of \lya optical depth around quasars.

The left panel of Figure~\ref{fig:sub_xvers_r3d} shows the inferred gas overdensity profiles for the transverse-distance subsamples, plotted as a function of three-dimensional separation $r_{\rm 3D}$ with a bin width of $0.33\,h^{-1}\mathrm{Mpc}$. These overdensities are derived from the $\tau_{\rm prx}/\tau_{\rm IGM}$ profiles using Equation~\ref{eq:tau_ratio}, with (solid lines) and without (dashed lines) the contribution of quasar ionizing radiation. Across most scales, all three subsamples agree well with one another in both cases with and without radiation contribution, except that the innermost subsample (black line) shows a marginally higher overdensity at $r_{\rm 3D} \sim 1{\rm -}5\hinvmpc$. If both the gas overdensity and quasar emission are isotropic, 
the three subsamples would be expected to yield identical profiles when binned by $r_{\rm 3D}$. In this sense, the overall consistency suggests either that (1) quasar emission is isotropic (corresponding to solid lines), or (2) the ionizing radiation cones have narrow opening angles such that their contribution to transverse sightlines is minimal (corresponding to dashed lines). However, the possibility of intermediate opening angles cannot be firmly excluded, as the stacking is performed over quasars whose emission axes are randomly oriented.

The right panels of Figure~\ref{fig:sub_all} all suggest some asymmetry in the transmission spectra relative to the location of the foreground quasar, such that there is more transmission on the far side than the near side. This asymmetry could be produced by a decrease in the neutral gas fraction on the far side relative to the near side, which is in the opposite sense of what would be expected from a finite lifetime. However, it is worth noting that residual uncertainties in quasar redshifts can also contribute to such apparent asymmetries: while the systematic bias has been corrected in DESI DR1 \citep{2025JCAP...01..130B}, scatter remains and may lead to asymmetric distributions when stacking large samples. 

To further investigate potential anisotropies in the quasar emission or environmental asymmetries, we examine the near-side versus far-side variation in the same transverse subsamples. The right panel of Figure~\ref{fig:sub_xvers_r3d} displays the ratio of the same profiles between pixels in the near side ($r_{\parallel} < 0$) and far side ($r_{\parallel} > 0$) relative to the foreground quasars. The ratios for all subsamples are consistent with unity, and no significant deviations are observed.
In other words, while the asymmetry is visually noticeable in the transmission spectra, it is not statistically significant when quantified through the near–far ratio. A formal two-sample KS test would provide one way to check this more rigorously, but the current results already suggest that any anisotropy or finite-lifetime signatures are weak compared to the statistical uncertainties or washed out in the stacked sample.

In particular, a significant contribution from finite quasar lifetimes would likely induce an asymmetric signal because the near and far sides correspond to different light-travel delays relative to the foreground quasar. This effect is expected to be strongest when the quasar lifetime is shorter than, or comparable to, the characteristic light-travel time to the transverse sightlines. 
The observed symmetry thus implies that, if the ionizing radiation is effectively isotropic, the quasar lifetime must be sufficiently long for the radiation to reach both sides of the foreground quasar, yielding an apparent near–far symmetry.
In this context, a characteristic scale is the transverse distance within which quasar ionization dominates over the UV background, i.e., where $\Gamma_q>\Gamma_{\rm UVB}$, or equivalently where $[\tau_{\rm prx}/\tau_{\rm IGM}]_{\rm rad}<0.5$. In Figure~\ref{fig:tau_ratio_luminosity}, for the middle luminosity bin, this criterion is satisfied only for the innermost transverse bin, with center $r_\perp = 1.5\,h^{-1}\,{\rm Mpc}$. Under this simple isotropic interpretation, the corresponding rest-frame light-travel time, of order a few Myr, can therefore be regarded as an illustrative lower-limit scale for the quasar lifetime.
In an earlier study based on a much smaller sample of quasar pairs, \citet{2008MNRAS.391.1457K} found no clear evidence for the transverse proximity effect in over 100 quasar pairs, and even reported enhanced absorption behind the foreground quasars, which they interpreted as evidence for a Myr-scale episodic lifetime. 

On the other hand, strongly anisotropic emission would reduce such lifetime constraints, since regions outside the cones may never be illuminated.
Moreover, if the ionizing radiation is emitted through bi-polar cones, the lack of near–far asymmetry could indicate that the orientation of the emission axes is randomly distributed and sufficiently averaged out in the stacking. This interpretation aligns with the earlier result that the transverse subsamples exhibit only marginal differences in the overdensity profiles, supporting the idea that quasar emission is either narrowly beamed or appears isotropic due to random orientations of the emission axes.

Another possibility is that the slightly enhanced overdensity seen in the innermost transverse subsample at $r_{\rm 3D} \sim 1{\rm -}5\hinvmpc$ may arise from redshift-space distortions (RSD), particularly the Fingers-of-God (FoG) effect. Although FoG primarily affects the line-of-sight direction, the innermost $r_\perp$ bin corresponds to scales comparable to typical halo sizes, where RSD can have a stronger influence. At fixed $r_{\rm 3D}$, sightlines with smaller transverse separations are more susceptible to line-of-sight velocity dispersion, which could artificially boost the inferred gas overdensity in these regions. This effect may offer an alternative explanation that is consistent with the modest excess seen only in the innermost bin and not accompanied by any significant near–far asymmetry.

\section{Summary} \label{sec:summary}

In this work, we have carried out a large-sample study of the TPE using more than 10,000 quasar pairs from the DESI Year 1 data. By analyzing the \lya absorption in the vicinity of foreground quasars, we examined how quasar radiation and environmental effects shape the surrounding IGM. Our analysis combined statistical measurements of transmission and optical depth profiles with model predictions of quasar ionization, allowing us to extract information on both gas overdensities and quasar emission properties. Our main findings are as follows:

\begin{itemize}
    \item \textbf{Dense environments dominate radiation:}\\
    The observed optical depth near quasars is enhanced relative to the IGM average, consistent with strong clustering of \ion{H}{1} absorbers. This interpretation is supported by the inferred gas overdensity profiles, which remain high at scales $\lesssim 10\hinvmpc$, especially around luminous quasars.  
    
    \item \textbf{Luminosity subsamples \newline \hspace*{7em} -- no clear dependence:}\\
    The measured \lya optical depth shows no clear dependence on foreground quasar luminosity. This lack of luminosity dependence could arise from an almost coincidental balance between higher ionizing flux and higher gas overdensity. Explaining it purely by environment, however, would require strong luminosity-dependent clustering, in conflict with clustering studies. Alternatively, it may reflect emission that is time-variable or anisotropic such that the ionizing radiation emitted toward us does not represent that toward other directions.
    
    \item \textbf{Transverse-distance subsamples \newline \hspace*{7em} -- apparent isotropy:}\\
    The stacked transverse-distance subsamples yield nearly identical overdensity profiles. This apparent isotropy may reflect either intrinsically isotropic emission or moderate-angle cones averaged out by random orientations. Alternatively, strongly anisotropic emission with narrow cones could simply miss most transverse sightlines.
    
    \item \textbf{Constraints from near–far symmetry:}\\
    The near–far ratio shows no significant difference, implying that if emission is isotropic, quasar lifetimes must be long enough for radiation to reach both sides (suggesting an illustrative lower-limit scale of order a few Myr for the quasar lifetime). The lack of asymmetry is also consistent with anisotropic emission, either with narrow cones that miss the transverse sightlines or with moderate cones whose orientations average out in the stacked sample.
    
    \item \textbf{Comparison with previous TPE studies:}
    Earlier work at $z=2{\rm -}4$ \citep[e.g.,][]{2004ApJ...610..642C,2008MNRAS.391.1457K,2013ApJ...776..136P,2019ApJ...884..151J} similarly reported excess \ion{H}{1} absorption in the transverse direction, often interpreted as signatures of dense environments and/or anisotropic quasar emission. Our results provide a consistent picture with these smaller-sample studies, now confirmed with unprecedented statistical power.
\end{itemize}  

A key challenge is to disentangle the combined effects of quasar emission properties (e.g., variation, lifetime, and geometry) and overdense environments. \citet{2011ApJ...736...49L} demonstrated that apparently contradictory lifetime estimates from previous proximity effect measurements may be reconciled when density enhancements and obscuration are modeled simultaneously. More recently, \citet{2025arXiv250403848C} developed a halo-model framework of the quasar proximity effect, calibrated and tested with hydrodynamic simulations, which can connect the proximity effect to both quasar host halo mass and the UVB on large scales. These studies highlight the importance of forward modeling with radiative transfer and realistic environments to fully exploit the statistical power of DESI.

Ongoing observations with DESI will significantly increase the size and quality of the dataset that may be used for future studies of the quasar proximity effect. The DESI collaboration is in the midst of analyzing the data that will form its second data release, which includes over 820,000 spectra that include the \lya forest \citep{2025arXiv250314739D}, and which have already been used to study the expansion history \citep{2025arXiv250314738D}. These new observations will increase the total number of quasars, increase the S/N of existing spectra, and increase the number of close pairs. These improvements will also facilitate more targeted tests of environmental absorption, for example by repeating the stacking analysis after splitting the sample by the presence/absence of LLSs at the foreground-quasar redshift once a robust DESI-based LLS catalog becomes available.

At higher redshifts, recent work has begun to map quasar light echoes and anisotropy directly. For example, \citet{2025arXiv250905417E} used JWST spectroscopy of background galaxies for a foreground quasar at $z\sim6$ to reveal the first tomographic detection of its light echo, constraining the orientation and lifetime of a luminous quasar. At lower redshift, \citet{2022ApJ...933..239M} examined the transverse proximity effect around BAL quasars, which are thought to be viewed from near the equatorial plane of the AGN torus and thus provide a probe of transverse emission. Their small sample did not show statistically significant differences from non-BAL quasars. 
In future DESI analyses, it will be interesting to compare the transverse proximity effect around BAL and non-BAL quasars, which could provide further insights into the anisotropy of quasar emission. 
External estimates of the obscured quasar fraction from X-ray or mid-infrared (MIR) observations may also provide an independent consistency check on anisotropic-emission interpretations and potentially an additional constraint on the opening angle, since in a simple geometric picture a larger obscured fraction corresponds to a smaller unobscured solid angle.
Together, these studies show how anisotropy, obscuration, and finite lifetimes can be probed across cosmic time, and they provide complementary perspectives to the DESI measurements at $z\sim2{\rm -}3$. 

A powerful next step is to combine transverse and line-of-sight proximity effect measurements. As demonstrated by \citet{2019ApJ...884..151J}, joint analyses of the TPE and LPE provide decisive constraints on anisotropy. A major challenge for LPE studies is accurate continuum fitting in the \lya forest region, since the measurement relies on the immediate blue side of the quasar’s Ly$\alpha$ emission line. Recently, \citet{2024ApJ...976..143T} introduced LyCAN, a convolutional neural network trained on synthetic spectra derived from HST/COS data together with DESI mock spectra, which predicts the unabsorbed continuum in the forest region using only the red side of the \lya emission line. This approach achieves errors of only a few percent, making it especially promising for combining TPE and LPE measurements.


\section*{Acknowledgments}

RH would like to thank Matthew~Pieri and Siwei~Zou for their support as the Absorber topical-group chairs throughout this project. We are also grateful to J.\ Xavier~Prochaska and Naim~G\"oksel~Kara\c{c}ayl{\i} for their insightful discussions, and to Molly~Wolfson and Zhiwei~Pan for their helpful comments on the draft.
PM acknowledges support from the United States Department of Energy, Office of High Energy Physics under Award Number DE-SC-0011726. ZZ was supported by National Science Foundation (NSF) grant AST-2007499.

This material is based upon work supported by the U.S. Department of Energy (DOE), Office of Science, Office of High-Energy Physics, under Contract No. DE–AC02–05CH11231, and by the National Energy Research Scientific Computing Center, a DOE Office of Science User Facility under the same contract. Additional support for DESI was provided by the U.S. National Science Foundation (NSF), Division of Astronomical Sciences under Contract No. AST-0950945 to the NSF’s National Optical-Infrared Astronomy Research Laboratory; the Science and Technology Facilities Council of the United Kingdom; the Gordon and Betty Moore Foundation; the Heising-Simons Foundation; the French Alternative Energies and Atomic Energy Commission (CEA); the National Council of Humanities, Science and Technology of Mexico (CONAHCYT); the Ministry of Science, Innovation and Universities of Spain (MICIU/AEI/10.13039/501100011033), and by the DESI Member Institutions: \url{https://www.desi.lbl.gov/collaborating-institutions}. Any opinions, findings, and conclusions or recommendations expressed in this material are those of the author(s) and do not necessarily reflect the views of the U. S. National Science Foundation, the U. S. Department of Energy, or any of the listed funding agencies.

The authors are honored to be permitted to conduct scientific research on I'oligam Du'ag (Kitt Peak), a mountain with particular significance to the Tohono O’odham Nation.


\section*{DATA AVAILABILITY}

The numerical data underlying the figures presented in this paper, together with the quasar-pair catalog used in the analysis, are publicly available on Zenodo at \url{https://doi.org/10.5281/zenodo.19281859}. The parent quasar and \lya-forest data products are based on the public DESI DR1 data release.

%

\vspace{5mm}
\facility{
            Mayall (DESI).
            }


\software{
          NumPy \citep{2020Natur.585..357H},
          Scipy \citep{2020NatMe..17..261V},
          Matplotlib \citep{2007CSE.....9...90H},
          Astropy \citep{2013A&A...558A..33A,2018AJ....156..123A,2022ApJ...935..167A},
          healpy \citep{2019JOSS....4.1298Z},
          Statsmodels \citep{seabold2010statsmodels},
          Numba \citep{2015llvm.confE...1L},
          CAMB \citep{2000ApJ...538..473L}.
          }



\appendix

\section{Comparison of error estimation methods} \label{error_comparison}

\begin{figure}[h!]
\centering
    \includegraphics[width=0.5\textwidth]{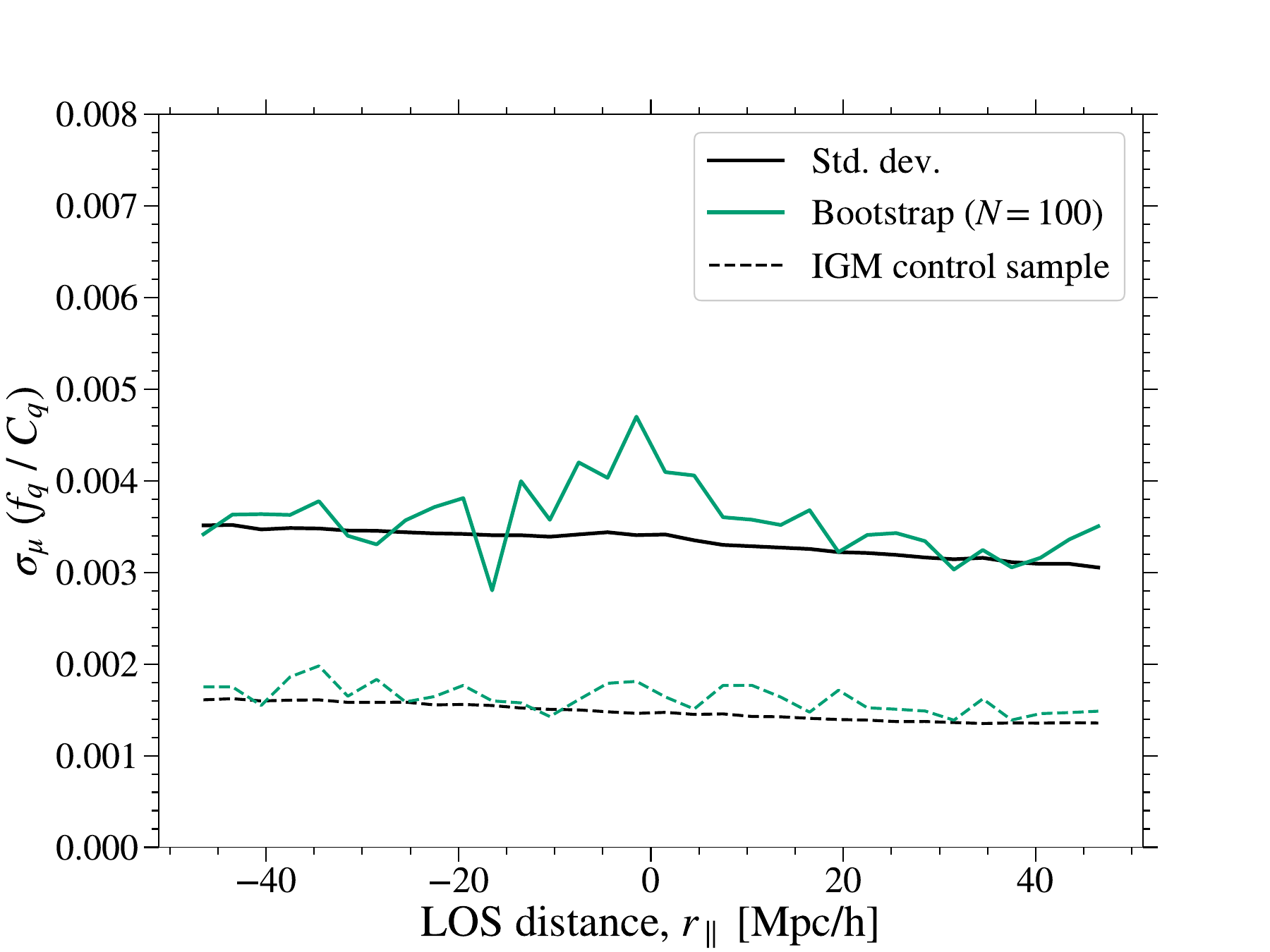}
\caption{Comparison of \lya transmission uncertainties estimated using two methods: the standard deviation across the entire sample (black) and bootstrap resampling with $N=100$ (green). Results are shown for both background quasars (solid lines) and the control samples (dashed lines).
\label{fig:std_vs_bootstrap}}
\end{figure}

In this paper, we adopted two different approaches to estimate the uncertainties in our measurements: the standard deviation for Ly$\alpha$ transmission and bootstrapping for other observables such as Ly$\alpha$ optical depth and gas overdensity. Here, we assess the consistency between these two methods (see the main text for details). Figure~\ref{fig:std_vs_bootstrap} shows the mean uncertainties of Ly$\alpha$ transmission, $\sigma_{\mu}(f_{q}/C_{q})$, computed for the entire sample of 10,942 background quasars (solid lines) and the control samples (dashed lines) using both the standard deviation and bootstrap resampling with $N=100$.

The two methods yield consistent results across most distances, with only a small deviation around $r_{\parallel}=0$. This confirms that our bootstrap approach provides sufficient precision for comparing different subsamples and for confronting the data with model predictions. Given that high-precision error estimation is not critical for our main scientific goals, we adopt the bootstrap method to estimate the mean and its uncertainty for all subsequent measurements in this study.

\bibliography{ms}{}
\bibliographystyle{aasjournal}



\end{document}